\documentclass{aa}  
\usepackage{graphicx}
\usepackage{txfonts}
\usepackage{comment}
\usepackage{color}
\usepackage{xcolor}
\usepackage{url}
\usepackage{hyperref}
\usepackage{multirow}
\usepackage{float}
\usepackage{orcidlink}
\usepackage{soul}
\usepackage{lipsum} 

\definecolor{myblue}{HTML}{6a359c}
\definecolor{myviolet}{HTML}{756BB1}
\definecolor{mygreen}{HTML}{7FCDBB}
\definecolor{myyellow}{HTML}{FFFDD1}

\hypersetup{
	colorlinks,
	linkcolor={myblue},
	citecolor={myblue},
	urlcolor={myblue}}
	
\begin{document} 

\title {A formation pathway for giant planets in S-type discs of $\gamma$-Cephei-like compact binaries}

 \titlerunning{}
\authorrunning{Ronco et al.}

\author{
M. P. Ronco\orcidlink{0000-0003-1385-0373}\inst{1}, 
O. M. Guilera\orcidlink{0000-0001-8577-9532}\inst{1},
J. Venturini\orcidlink{0000-0001-9527-2903}
\inst{2},
F. A. Zoppetti\orcidlink{0000-0002-3876-0887}\inst{3},
M. M. Miller Bertolami\orcidlink{0000-0001-8031-1957}\inst{1}
}
\offprints{mpronco@fcaglp.unlp.edu.ar}
\institute{
Instituto de Astrof\'{\i}sica de La Plata, CCT La Plata-CONICET-UNLP, Paseo del Bosque S/N (1900), La Plata, Argentina,
\and
Department of Astronomy, University of Geneva, Chemin Pegasi 51, 1290 Versoix, Switzerland
\and
Instituto de Astronom\'{\i}a Te\'orica y Experimental (CONICET-UNC). Laprida 854, X5000BGR C\'ordoba, Argentina
}

\date{}

\abstract{Planet formation in close binary systems such as $\gamma$-Cephei is strongly challenged by the severe truncation of the circumprimary disc induced by the stellar companion, which drastically limits the available reservoir of gas and solids. Recent hydrodynamical studies suggest that a long-lived circumbinary disc may replenish the circumprimary disc with gas and dust, extending its lifetime and potentially enabling giant planet formation. However, the long-term evolution of such systems under the combined effects of viscous accretion and X-ray photoevaporation, and their coupling with planet formation, remains largely unexplored.}
{We aim to investigate whether sustained mass inflow from a circumbinary reservoir can prolong the lifetime of circumprimary discs and facilitate gas giant planet formation in $\gamma$ Cephei–like binary systems, even in the presence of strong photoevaporative winds.}
{Using our code PLANETALP-B, we model the coupled evolution of gas, dust growth, and in-situ planet formation by pebble and gas accretion in a $\gamma$-Cephei-like circumprimary disc, including viscous accretion, X-ray photoevaporation, and continuous mass injection from an external circumbinary disc.}
{Gas inflow from the circumbinary disc can significantly extend the lifetime of the circumprimary disc, even under strong photoevaporative mass loss. When a fraction of solids is transferred, the lifetime of the circumprimary solid disc increases as well, enhancing the efficiency of planetary growth. As a result, our simulated planets can reach up to several Jupiter masses, in contrast to scenarios that neglect mass replenishment.}
{We show that sustained mass transfer from a circumbinary disc can indeed play a key role in enabling giant planet formation in $\gamma$-Cephei–like close binaries. This mechanism provides a viable pathway to overcome the limitations of disc truncation, although its applicability to other types of binary systems remains to be tested with dedicated hydrodynamical simulations.} 


\maketitle
%
\section{Introduction}


Stellar multiplicity is a common outcome of star formation, with roughly half of Sun-like stars residing in binary or higher-order stellar systems \citep{Raghavan2010,Offner2023}. While most planet-hosting binaries are found at wide separations of several hundred au \citep{Lester2021,ThebaultBonanni2025}, a small but growing number of systems are known in which planets orbit one component of a close or intermediate binary, with stellar separations of only a few tens of au. These systems are particularly intriguing because this is precisely the regime where dynamical perturbations from the stellar companion are expected to have the strongest impact on protoplanetary discs and on the planet formation process itself.
Although observational statistics in this separation range remain limited \citep[e.g.,][]{MarzariThebault2019}, the existence of such systems poses a significant theoretical challenge: planets are observed where planet formation should, in principle, be most strongly inhibited. As the number of known S-type planets continues to increase, driven by surveys such as Gaia \citep{Mugrauer2021, Behmard2022, Mugrauer2023} and high-resolution imaging \citep{Lester2021, Sullivan2023}, it becomes essential not only to understand S-type planet formation in general, but in particular how such planets can form and survive in these more compact and dynamically hostile binary systems.

From a theoretical perspective, circumstellar discs in binary systems are subject to strong tidal interactions from the companion star, which truncate and heat the disc \citep{PapaloizouPringle1977, ArtymowiczLubow1994}. This truncation can severely limit planet formation by reducing the amount of gas and dust available, an effect that is particularly pronounced in close binary systems \citep{RosottiClarke2018, Zagaria2021} and has been confirmed through disc observations with ALMA (\citealp[see][and references there in]{Zurlo2023}). In addition, gravitational perturbations from the companion excite the eccentricities and inclinations of growing planetary embryos, with increasingly disruptive effects as the binary separation decreases \citep{MarzariThebault2019} and as the planet location aproaches the truncation radius \citep{Nigioni2026}. Nevertheless, roughly 40 close binary systems with separations smaller than $\sim$30 au are known to host planets, many of them gas giants, posing a serious challenge to standard planet formation models (\citealp[see][]{ThebaultBonanni2025}, and Table 1 in \citealp{MarzariDangelo2025}).

Among these systems, $\gamma$ Cephei stands out as one of the most studied and representative examples. It consists of a binary system with a primary star of 1.4$~M_\odot$ and a secondary star of 0.4$~M_\odot$ \citep{Neuhauser2007}, separated by $\sim$20 au with an eccentricity of $e \sim 0.4$ \citep{Endl2011}, and hosting a giant planet, the first one detected in a close-in binary system, with a mass of $\sim$1.6$~M_{\rm Jup}$ orbiting at $\sim$2 au \citep{Hatzes2003}. More recently, an asteroseismic analysis by \citet{Knudstrup2023} revisited the system parameters, resulting in $M_1 \sim $1.27$~M_\odot$, $M_2 \sim$ 0.328$~M_\odot$, and a planetary mass of $\sim$6.6$~M_{\rm Jup}$. Although the planet is currently observed on an orbit nearly perpendicular to the binary plane, this configuration is generally interpreted as the result of post-formation dynamical processes, such as a stellar fly-by \citep{MartiBeauge2012} or the eccentric Kozai–Lidov mechanism \citep{HuangJi2022}.

Forming the giant planet in $\gamma$ Cephei could be particularly challenging. The main difficulty stems from the strong truncation of the circumprimary/circumsecondary disc by the eccentric stellar companion, which severely limits the gas and solid reservoir available for planet growth \citep{MullerKley2012,RosottiClarke2018,Zagaria2021}. Hydrodynamical studies further show that the disc can become eccentric and start precessing \citep{Paardekooper2008}, and together with stellar perturbations this may inhibit planetesimal accretion by increasing their impact velocities beyond the fragmentation threshold \citep{Thebault2004}. Although gas drag can partially reduce impact velocities through orbital alignment, this mechanism is only effective under restrictive conditions, such as only considering planetesimals of the same sizes \citep{MarzariScholl2000}, or during the specific phase of disc dissipation \citep{Xie2008}. Also, disc precession may partially mitigate this effect if the discs are highly eccentric \citep{Beauge2010}. In this context, more detailed secular analyses have shown that classical first-order approximations may significantly misestimate key dynamical quantities such as the forced eccentricity and precession frequency, affecting planetesimal accretion \citep{Giuppone2011}.
 More recently, \citet{Camargo2023,Camargo2024} used hydrodynamic simulations of truncated circumprimary discs to show that a giant planet can form in situ or migrate to the orbit of $\gamma$-Cephei b. However, their models assume a massive seed ($\sim$0.1 M$_\text{J}$), bypassing the challenging early stages of core formation in such close binaries.
 
Together, these results suggest that the main bottleneck for planet formation in this system is the early growth of a massive core. This motivates alternative pathways, such as pebble accretion, which may be more efficient in truncated and dynamically perturbed environments.

In this line, \citet{Venturini2026} and \citet{Nigioni2026} recently presented the PAIRS project, the first global planet formation model for S-type planets within the pebble accretion paradigm. They showed that, despite disc truncation limiting the pebble flux, gas giants ($>100 $M$_\odot$) can form for binary separations $\gtrsim 40$ au and truncation radii $\gtrsim 7$ au. However, reproducing $\gamma$-Cephei–like planets remains challenging within their framework. 

A possible solution to the problem of forming $\gamma$-Cephei b has recently been explored by \citet{MarzariDangelo2025}, who adopted a scenario in which a long-lived circumbinary disc, remnant of the star formation process, feeds the circumprimary disc with gas and solids \citep{Dutrey1994,NelsonMarzari2016}. Using high-resolution hydrodynamical simulations including gas and dust evolution, they tested this scenario and showed that, in $\gamma$ Cephei–like binaries, mass transfer can extend the circumprimary disc lifetime to $\sim$3 Myr, roughly three times longer than in isolation. Moreover, solid particles are also delivered to the circumprimary disc, suggesting that disc replenishment could play a non-negligible role in planet formation in close binary systems. However, dust filtration at the tidal gap causes 1 mm grains to remain trapped outside the gap, leading to a non-continuous size distribution. For the specific parameters of $\gamma$-Cephei, they further found no mass transfer from the circumprimary to the circumsecondary disc, while the gas accreted by the secondary from the gap region is three orders of magnitude lower than that accreted by the primary, preventing a persistent disc around the secondary. While these results demonstrate that disc replenishment may substantially modify the gas and dust budget of circumprimary discs in close binaries, the implications of this mechanism for the actual formation of planets were not explored in that work.

To assess whether disc replenishment can facilitate giant planet formation, we study the long-term evolution of a circumprimary disc in a binary system that is fed by an external circumbinary disc, using an adapted version of the 1D+1D disc evolution model PLANETALP-B \citep{Ronco2021}. Our model includes viscous accretion, stellar irradiation, and X-ray photoevaporation from the primary/binary star, not considered in \citet{MarzariDangelo2025}. The model also includes dust growth and evolution, and is coupled to in-situ planet formation via pebble and gas accretion. This framework allows us to evaluate whether the extended disc lifetimes and enhanced solid reservoirs produced by mass injection can lead to the formation of gas giant planets. We find that this replenishment mechanism indeed enables the formation of $\gamma$-Cephei-like planets, offering a viable pathway to planet formation in compact binaries.

The paper is organized as follows. Section 2 presents a detailed description of our model, including the evolution of both the circumbinary and circumprimary discs, and the planet formation squeme, section 3 describes the initial condition setups, and section 4 presents the results. A detailed discussion concerning model caveats is presented in section 5 and the conclusions are resumed in section 6.

\section{Model Implementation}

PLANETALP-B is a 1D+1D code that computes the time evolution of a gaseous disc in different possible scenarios involving multiple star systems. The code, applicable to S-type (circumprimary/ circumsecondary) and P-type (circumbinary) protoplanetary discs, was originally developed to study disc evolution in hierarchical triple-star systems and applied to HD 98800 to explain the longevity of its disc \citep{Ronco2021}.

\begin{figure*}[ht]
   \begin{center}
   \includegraphics[width=0.85\linewidth]{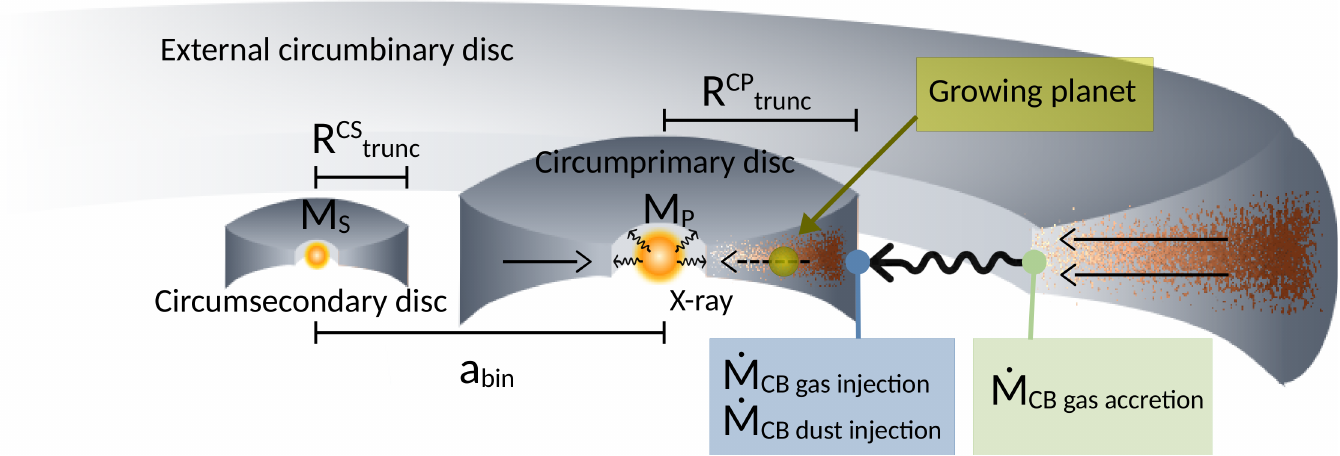}
   \caption{Schematic view of the configuration of our scenario of study: an axisymmetric circumbinary (also called P-type) disc around the close-in binary star system, with circumprimary/circumsecondary (also called S-type) discs around each stellar component.}
   \label{Fig1}%
   \end{center}
   \end{figure*}

Here we adapt PLANETALP-B to compute the time evolution of the gas and dust components of a circumprimary disc (CP hereafter) in a binary star system that is fed with gas and dust from an external circumbinary disc (CB hereafter), remnant from the star formation process.
A schematic view of the configuration of this case can be seen in figure \ref{Fig1}. We also compute the in-situ planet formation adopting the formation module of PLANETALP (the single star planet formation version) described in \citet{Ronco2017,Guilera2017, Guilera2020,Venturini2020Letter, Venturini2020SE}. We apply our model to a $\gamma$-Cephei-like system, with the aim of testing if 
extra gas and solids comming from an external CB disc can ideed help forming gas giant planets.

For clarity, the main steps undertaken to achieve our goal, which are discussed in detail in the following subsections, are:
\begin{enumerate}
\item We first compute the time evolution of the gaseous component of the CB disc and record the mass accretion rate onto the inner binary. Following the results of \citet{MarzariDangelo2025}, we assume this accretion is transferred mainly to the CP disc through tidal streams (\citealp[see][for details]{Ronco2021}), and discard mass transfer to the circumsecondary (CS hereafter) disc. We consider models with and without CB disc photoevaporation.

\item We then compute the gas and dust evolution of the CP disc, which is fed by the previously determined accretion rate. A fixed fraction of this gas inflow is assigned to the dust component. The CP disc is always affected by photoevaporation from the central star.

\item Finally, we model the coupled evolution of gas, dust, and in-situ planet formation at three distinct locations within the CP disc. This allows us to assess whether the additional material supplied by the CB disc can promote the formation of gas giant planets.

\end{enumerate}

\subsection{The vertical structure of the CB and CP discs}

Due to the axisymmetrical nature of our 1D+1D model, we consider circular and coplanar orbits for the binary star system, the CP, CS, and CB discs.  
Moreover, while the CP disc, truncated by the stellar companion, rotates around the primary star, the CB disc, which has an inner cavity generated by the torques of the inner binary, is considered to be rotating in a gravitational potential given by the sum of both stellar masses. 

To compute the vertical structure of the CB and CP discs, we assume axi-symmetric and irradiated discs in hydrostatic equilibrium. 
Adapting the equations of Sec. 2.1 of \citet{Ronco2021} to our case of study, the complete set of equations is given by:
 \begin{eqnarray}
  \frac{\partial P}{\partial z} &=& -\rho \Omega^2 z, \label{eq:Presion}\\
    {\frac{\partial F}{\partial z}} &=& \frac{9}{4} \rho \nu \Omega^2 + D_{\Lambda}, \label{eq:Flujo} \\
    \frac{\partial T}{\partial z} &=& \nabla \frac{T}{P}\frac{\partial P}{\partial z},
 \label{eq:Temperatura} 
\end{eqnarray}
where $P$, $\rho$, $F$, $T$ and $z$ represent the pressure, density, radiative heat flux, temperature, and vertical coordinate of the disc, respectively. $\nu= \alpha c_s^2/\Omega$ is the viscosity \citep{Shakura1973}, where $\alpha$ is a dimensionless parameter. $c_s^2= P/\rho$ is the square of the locally isothermal sound speed, and $\Omega$ is the Keplerian frequency at a given radial distance $R$ from the central binary or primary,
\begin{equation}
\Omega = 
\begin{cases}
\sqrt{GM_{\text{B}}/R^3}, & \text{for the CB disc}\\
\sqrt{GM_{\text{P}}/R^3}, & \text{for the CP disc},
\end{cases}
\label{eq_frecuencia}
\end{equation}
where $M_{\text{B}}=M_{\text{P}}+M_{\text{S}}$ with $M_{\text{P}}$ the mass of the primary and $M_{\text{S}}$ the mass of the secondary. 

The term $D_{\Lambda}$ in Eq. \ref{eq:Flujo} represents the tidal heating of the CB or CP disc, depending on the case, dissipated in the form of radiation, and is given by 
\begin{equation}
D_{\Lambda} = 
\begin{cases}
(\Omega_{\text{B}} - \Omega)\Lambda_{\text{CB}}\rho,\text{~~~for the CB disc}\\
(\Omega_{\text{B}} - \Omega)\Lambda_{\text{CP}}\rho, \text{~~~for the CP disc}
\end{cases}
\label{eq_gas-evol}
\end{equation}
where $\Omega_\text{B}$ is the Keplerian frequency binary separation given by $\Omega_\text{B}=\sqrt{\frac{GM_{\text{B}}}{a_{\text{B}}}}$, with $a_{\text{B}}$ the binary separation. $\Lambda_{\text{CB}}$ and $\Lambda_{\text{CP}}$ represent the torques generated by the binary on the CB and CP discs, respectively, following \citet{ArmitageNatarajan2002, TazzariLodato2015, Fontecilla2019} and \citet{Ronco2021}, given by 
\begin{eqnarray}
\Lambda_{\text{CB}}(R) &=& \frac{f}{2}q_{\text{B}}^2\Omega^2R^2\left(\frac{a_{\text{B}}}{\Delta_{\text{B}}}\right)^4 e^{-\left(\frac{R-R_{\text{OM}}}{W_{\text{OM}}}\right)^2}, \label{eq:TorqueI} \\
\nonumber \\
\Lambda_{\text{CP}}(R) &=& -\frac{f}{2}q_{\text{B}}^2\Omega^2R^2\left(\frac{R}{\Delta_{\text{B}}}\right)^4 e^{-\left(\frac{R-R_{\text{IM}}}{W_{\text{IM}}}\right)^2},
\label{eq:TorqueII}
\end{eqnarray}
where $f$ is a dimensionless normalization parameter between 0.001 and 1, \citep{ArmitageNatarajan2002, Alexander2012, Vartanyan2016, Shadmehri2018} taken as 0.01, $q_{\text{B}}$ is the mass ratio between $M_{\text{S}}$ and $M_{\text{P}}$, and $\Delta_{\text{B}} = {\text{max}}(R^{\text{B}}_{\text{Hill}},{\text{H}}_{\text{g}},|R-a_{\text{B}}|)$  
with $R^{\text{B}}_{\text{Hill}}=a_{\text{B}}(q_{\text{B}}/3)^{1/3}$ the Hill radius of the secondary star in the binary system, and ${\text{H}}_{\text{g}}$ is the height scale of the disc. As in \citet{Fontecilla2019} we consider $R_{\text{OM}}=1.59a_{\text{B}}$ and $R_{\text{IM}}=0.63a_{\text{B}}$ being the radii of the outermost and innermost Lindblad resonances,  and $W_{\text{OM}}=75{\text{H}}_{\text{g}}$ and $W_{\text{IM}}=370{\text{H}}_{\text{g}}$ being the widths of the Gaussian smoothing. \\

For disc irradiation, we follow Eqs. 6–10 of \citet{Ronco2021}, assuming that the CB disc is irradiated by both stars, while the CP disc is irradiated only by the primary. This is justified by the negligible luminosity contribution of the secondary at low mass ratios \citep{Venturini2026}.

Finally, we note that solving these equations provides the mean viscosity $\overline{\nu}(\Sigma_{\text{g}}, R)$, needed to model the time evolution of the gas disc. 

\subsection{The CB gas disc evolution}

We first compute the time evolution of the gas surface density of the CB disc. We solve the classical 1D diffusion equation \citep{Pringle1981}, including the torque term due to the binary system, previously described (Eq.\ref{eq:TorqueI}), which generates an inner cavity between the binary and the CB disc,
\begin{align}
  \frac{\partial \Sigma^{\text{CB}}_{\text{g}}}{\partial t}= & \frac{3}{R}\frac{\partial}{\partial R} \left[ R^{1/2} \frac{\partial}{\partial R} \left( \overline{\nu} \Sigma^{\text{CB}}_{\text{g}} R^{1/2}  \right) - \frac{2\Sigma \Lambda_{\text{CB}}}{3\Omega} \right] - \dot{\Sigma}^{\text{CB}}_{\text{w}}(R), 
\label{eq:evol_gas_CB}
\end{align}
where $t$ represents time, $\overline{\nu}$ the mean viscosity at the mid-plane of the disc, $\Sigma^{\text{CB}}_{\text{g}}$ the gas surface density and $\dot{\Sigma}^{\text{CB}}_{\text{w}}(R)$ the sink term due to the X-ray photoevaporation computed following the analytical prescriptions derived by \citet{Owen2012} (see their Apendix B) and the implementation in \citet{Ronco2021} (see their Sec. 2.2) which yield characteristic mass loss rates of  $\dot{M}\sim10^{-8}~$M$_\odot$ yr$^{-1}$. It is worth noting that the photoevaporation of the CB disc could be mitigated by the presence of both the CP and CS discs, which may shield the external disc from direct stellar radiation. Given the uncertainty in the efficiency of this effect, we therefore perform simulations including and excluding CB disc photoevaporation, treating these two configurations as limiting cases.

As explained in \citet{Ronco2021}, due to the 1D nature of our approach, gas accretion onto the binary system is suppressed by construction. However, 2D/3D simulations show it can occur via tidal streams, typically at 10–60\% of the standard rate \citep{ArtymowiczLubow1996,Cuadra2009, Farris2014,Dunhill2015, Tang2017}. In \citet{Ronco2021} (see their Appendix B) we showed that considering an accretion efficiency of 50$\%$ does in fact matches the more realistic approach derived by \citet{Ragusa2016} who considered an efficiency that depends on the disc aspect ratio. Thus, we allow a fraction 50$\%$ of the expected rate to be accreted by the binary. This accretion rate is computed as $\dot{M}_{\text{CB}}=3\pi\nu\Sigma^{\text{CB}}_{\text{g}}$ at every radial bin between $R^{\text{CB}}_{\text{cav}}$ and $2R^{\text{CB}}_{\text{cav}}$, where $R^{\text{CB}}_{\text{cav}}$ is the radius of the inner cavity generated by the binary torques, and adopt the maximum value of $\dot{M}_{\text{CB}}$ in that region, considering it as an upper limit.

Equation \ref{eq:evol_gas_CB} is solved using an implicit Crank–Nicholson scheme over 1000 logarithmically spaced radial bins between $a_{\text{B}}$ and 1000 au. At these positions, we set zero-torque boundary conditions, which is equivalent to having zero density at the boundaries. However, the tidal torque term in Eq. \ref{eq:evol_gas_CB} naturally pushes the disc away from the binary forming a cavity between them (\citealp [see][for details]{Ronco2021}).

\subsection{The CP gas and dust disc evolution}

Once we have computed the CB disc evolution and registered $\dot{M}_{\text{CB}}$ at every time step, we can calculate the  evolution of the gas surface density of the CP disc by solving:
\begin{align}
  \frac{\partial \Sigma^{\text{CP}}_{\text{g}}}{\partial t}= & \frac{3}{R}\frac{\partial}{\partial R} \left[ R^{1/2} \frac{\partial}{\partial R} \left( \overline{\nu} \Sigma^{\text{CP}}_{\text{g}} R^{1/2}  \right) - \frac{2\Sigma \Lambda_{\text{CP}}}{3\Omega} \right] - \dot{\Sigma}^{\text{CP}}_{\text{w}}(R) + \dot{\Sigma}^{\text{CB}}_{\text{g}}(R),
\label{eq:evol_gas_CP}
\end{align}
where $\Sigma^{\text{CP}_{\text{g}}}$ is the gas surface density of the CP disc, disc$\dot{\Sigma}^{\text{CP}}_{\text{w}}(R)$ is the sink term due to the X-ray photoevaporation due to the primary star, and $\dot{\Sigma}^{\text{CB}}_{\text{g}}(R)$ is the source term that injects the gas lost by the CB disc through its cavity. Following \citet{MarzariDangelo2025}, who showed that for the parameters of $\gamma$ Cephei the fraction of CB mass intercepted by the secondary is negligible, we assume that all gas lost by the CB disc through tidal streams ($\dot{M}_{\text{CB}}$) is injected in the CP disc. At each timestep, we interpolate the mass inflow rate $\dot{M}_{\text{CB}}$
derived from the CB disc evolution. The injected mass is then distributed radially according to a Gaussian profile centered on the initial CP disc truncation radius $R^{\text{CP}}_{\text{trunc}}$
 (see Sec. \ref{IC}).

Equation \ref{eq:evol_gas_CP} is also solved over 1000 logarithmically equally spaced radial bins, but between an inner radius defined as $R_{\text{in}}=0.1$~au and $a_{\text{bin}}$. We also adopt zero-torque boundary conditions, with the tidal torque from the secondary star self-consistently setting the disc size to the truncation radius, $R^{\text{CP}}_{\text{trunc}}$.

To compute the growth and time evolution of the dust surface density we follow the same procedure described in \citet{Guilera2020} and \citet{Venturini2020SE}. The dust growth follows the approach of \citet{Drazkowska2016,Drazkowska2017}, based on the results of \citet{Birnstiel2011, Birnstiel2012}. The initial and minimum size of the dust particles at each radial bin in the CP disc is set to 1 $\mu$m, and the maximum is limited by dust coagulation, radial drift and dust fragmentation (see Eq. 8-12 in \citet{Guilera2020}. In this work, for simplicity, we consider the same fragmentation velocity threshold along the disc of $10$ m~s$^{-1}$, valid for icy grains \citep{GundlachBlum2015}, as in \citet{Rosotti2019b} and \citet{Zagaria2021}. 

The time evolution of the dust surface density of the CP disc, $\Sigma^{\text{CP}}_{\text{d}}$, is computed by solving the advection-diffusion equation, 
\begin{equation}
  \frac{\partial}{\partial t} \left(\Sigma^{\text{CP}}_{\text{d}}\right) + \frac{1}{r} \frac{\partial}{\partial r} \left( r \,\overline{v}_{\text{drift}} \,\Sigma^{\text{CP}}_{\text{d}} \right) - \frac{1}{r} \frac{\partial}{\partial r} \left[ r \,D^{*} \,\Sigma^{\text{CP}}_{\text{d}} \frac{\partial}{\partial r} \left( \frac{\Sigma^{\text{CP}}_{\text{d}}}{\Sigma^{\text{CP}}_{\text{g}}} \right) \right] = \dot{\Sigma}^{\text{CB}}_{\text{d}}(R),  
  \label{eqAdvDif}
\end{equation}
where $D^{*}= \nu / ( 1 + \text{St}^2 )$ is the dust diffusivity \citep{YoudinLithwick2007}, $\overline{v}_{\text{drift}}$ is the weighted mean drift velocity of the pebbles population (see \citet{Guilera2020} for more details) and $\dot{\Sigma}^{\text{CB}}_{\text{d}}$is the dust source term from the CB disc. 

As the time evolution of the dust component in the CB disc is not modelled in this work, we simply assume that $\dot{\Sigma}^{\text{CB}}_{\text{d}}=Z\dot{\Sigma}^{\text{CB}}_{\text{g}}$, where $Z = Z_0 10^{[\text{Fe}/\text{H}]}$ is the dust-to-gas ratio, with $Z_0=0.0187$ the primordial abundance of heavy elements in the Sun \citep{Lodders2025} and $[\text{Fe}/\text{H}]$ the stellar metallicity. The dust injection is then modeled the same way as for the gas. The size of the injected dust particles is, for simplicity, the same size of the dust/pebbles of the CP disc at the injection position. The limitations of this assumption are discussed in sec. \ref{sec:discussion}. Additionally, we did not model pebble sublimation inside the snow line to keep the model as simple as possible.

It is worth noting that the dust evolution is not directly influenced by the binary dynamics (i.e. no explicit binary term appears in Eq. \ref{eqAdvDif}), but rather indirectly through the evolution of the gas surface density profile.

\subsection{Our planet formation model}
\label{sec:planet_formation_model}

We compute the in-situ growth of a single moon mass protoplanetary embryo, initially located at different positions within the CP disc, using a simplified version of the planet formation module of PLANETALP \citep{Guilera2020, Venturini2020SE,Venturini2020Letter}. 
The planet grows by pebble accretion \citep{LambrechtsJohansen2014,Brasser2017} until it reaches its pebble isolation mass \citep{Lambrechts+2014}, after which it continues growing by accreting the surrounding gas \citep{Ikoma2000,TanigawaIkoma2007}.
For simplicity, we neglect gravitational interactions between the planet and the secondary star. The companion’s effect is included only through the truncation of the gas and dust disc, while the planet is assumed to remain on a circular, coplanar orbit, as in \citet{Venturini2026}. The limitations of this assumption are discussed in Section \ref{sec:discussion}.

\section{Initial condition setups}
\label{IC}

In this section we describe the initial conditions adopted in our simulations. We consider two main setups: a fiducial case assuming a circular binary orbit, and a second, more realistic case that accounts for the binary eccentricity. In both cases, the truncation radii of the CP and CS discs, which determine the initial disc mass in each case, are computed using the open-source Python script provided by \citet{Venturini2026} \citet{Venturini2026} (\url{https://github.com/Aryy98/Circumstellar_disc_truncation_radius}). Since we focus on a $\gamma$ Cephei–like system, we adopt the same stellar masses as in \citet{MarzariDangelo2025}, determined by \citet{Neuhauser2007}, and consider a $[\text{Fe}/\text {H}]=0.2$ \citep{Knudstrup2023}.

We begin by assuming that the total initial gas mass in the system before disc truncation, $M^{\text{ini}}_{\text{sys}}$, is 10\% of the mass of the inner binary. This mass is distributed according to a single gas surface density profile, common to all discs, given by
\begin{eqnarray}
\Sigma{_\text{g}} &=& \Sigma_{\text{g}}^0 \left( \frac{R}{R_{\text{c}}} \right)^{-\gamma} e^{-(R/R_{\text{c}})^{2-\gamma}},
\end{eqnarray}
where $R_{\text{c}}$ is the characteristic radius, $\gamma$ the surface density exponent, and $\Sigma_{\text{g}}^0$ a normalization constant. For simplicity, we adopt $R_{\text{c}}=40$~au and $\gamma=1$ \citep{Andrews2009}.

The initial masses of the CP and CS discs are then obtained by truncating this same profile at their corresponding truncation radii and integrating $\Sigma_{\text{g}}$ between the inner radius $R_{\text{in}}=0.1$~au and the corresponding $R_{\text{trunc}}$.

The mass of the external CB disc is computed following a similar procedure, integrating the same surface density profile from an inner cavity whose size is given by the binary semimajor axis in the circular case, and by the binary apocentre in the eccentric case. In this way, the sum of the masses of the three discs corresponds to the initially settled disc configuration resulting from truncating a single primordial surface density distribution. We note that, after the beginning of the simulations, the inner cavity of the CB disc is further displaced outward to $R^{\text{CB}}_{\text{cav}}$ as a consequence of solving Eq.~\ref{eq:evol_gas_CB}.

As for the solids, the initial dust surface density profile for the CP disc in computed as $\Sigma^{\text{CP}}_{\text{d}}=Z{\Sigma}^{\text{CP}}_{\text{g}}$, where $Z=0.029$ is the dust-to-gas ratio derived from the stellar metallicity.

Table \ref{tab:1} shows the initial conditions for the circular and eccentric cases. In what follows, we will describe in detail both.

\begin{table}
    \centering
    \renewcommand{\arraystretch}{1.5} 
  \begin{tabular}{cccc}
  \hline 
  \hline
  \multicolumn{2}{c}{} & Circular Case & Eccentric Case \\
  \hline
  \multirow{9}{*}{\rotatebox{90}{Stellar Parameters}} & $M_{\text{P}}$ [$M_\odot$] & \multicolumn{2}{c}{1.40} \\
\cline{2-4}
& $M_{\text{S}}$ [$M_\odot$] & \multicolumn{2}{c}{0.40} \\
\cline{2-4}
& $T_{\text{P}}$ [K] & \multicolumn{2}{c}{4647} \\
\cline{2-4}
& $T_{\text{S}}$ [K] & \multicolumn{2}{c}{3670} \\
\cline{2-4}
& $R_{\text{P}}$ [$R_\odot$] & \multicolumn{2}{c}{3.601} \\
\cline{2-4}
& $R_{\text{S}}$ [$R_\odot$] & \multicolumn{2}{c}{2.338} \\
\cline{2-4}
& $q_{\text{B}}$ & \multicolumn{2}{c}{0.286} \\
\cline{2-4}
& $a_{\text{B}}$ [au] & \multicolumn{2}{c}{20}\\
\cline{2-4}
& $e_{\text{B}}$  & 0 & 0.4\\
\cline{1-4}
\multirow{9}{*}{\rotatebox{90}{Disc Parameters}}
& $\gamma$ & \multicolumn{2}{c}{1} \\
\cline{2-4}
& $R_{\text{c}}$ [\text{au}] & \multicolumn{2}{c}{40}\\
\cline{2-4}
& $M^{\text{ini}}_{\text{sys}}$ [$M_\odot$] &  \multicolumn{2}{c}{0.180} \\ 
\cline{2-4}
& $R^{\text{CP}}_{\text{trunc}} [\text{au}]$ & 8.49 &  4.53 \\ 
\cline{2-4}
& $M^{\text{CP}}_{\text{d}}$ [$M_\odot$] & 0.035 & 0.019\\ 
\cline{2-4}
& $R^{\text{CS}}_{\text{trunc}} [\text{au}]$ & 4.81 &  2.65 \\ 
\cline{2-4}
& $M^{\text{CS}}_{\text{d}}$ [$M_\odot$] & 0.019 & 0.010 \\
\cline{2-4}
& $R^{\text{CB}}_{\text{cav}} [\text{au}]$ & $\sim$40  & $\sim$54 \\ 
\cline{2-4}
& $M^{\text{CB}}_{\text{d}}$ [$M_\odot$] & 0.108 & 0.088 \\ 
\cline{1-4}
& $M^{\text{fin}}_{\text{sys}}$ [$M_\odot$] & 0.162 & 0.117     \\
\cline{1-4}
\end{tabular}
    \vspace{0.2cm}
    \caption{Stellar and protoplanetary disc initial conditions used in our simulations. Subscripts P (S) and CP (CS) denote the primary (secondary) star and its circumstellar disc, respectively, while CB refers to the CB disc. Stellar temperature $T$ and radius $R$, necessary to compute the disc vertical structure, are taken from \citet{Baraffe2015}. The binary parameters $q_{\text{B}}$, $a_{\text{B}}$ and $e_{\text{B}}$ correspond to the mass ratio, semimajor axis and eccentricity taken as in \citet{MarzariDangelo2025}. The disc properties $\gamma$, $R_{\text{c}}$ and $M^{\text{ini}}_{\text{sys}}$ denote the surface density exponent, characteristic radius, and total initial mass (prior to truncation). $R^{\text{CP}}_{\text{trunc}}$ and $R^{\text{CS}}_{\text{trunc}}$ are the CP and CS disc truncation radius. $R^{\text{CB}}_{\text{cav}}$ is the cavity size of the CB disc and a result of our simulations. The disc masses after truncation are $M^{\text{CP}}_{\text{d}}$, $M^{\text{CS}}_{\text{d}}$ and $M^{\text{CB}}_{\text{d}}$, and the final system mass is $M^{\text{fin}}_{\text{sys}}=M^{\text{CP}}_{\text{d}}+M^{\text{CS}}_{\text{d}}+M^{\text{CB}}_{\text{d}}$.}
    \label{tab:1}
\end{table}

\section{Results}

In the following sections, we describe in detail the main results of our simulations of disc evolution and planet formation for the circular and eccentric cases. We note that both, the CB and CP disc evolution simulations end when the discs completely dissipate, this is when their mass drops to $10^{-6}M_\odot$, or after 5 Myr of evolution, whichever occurs first. Note that it is possible, that the CP disc dissipates faster than the CB disc. 

\subsection{The circular case}
\subsubsection{The CB disc evolution}
\label{sec:CBdiscevolution}
\begin{figure*}[ht!]
  \centering
    \includegraphics[angle= 0, width= 0.9\textwidth]{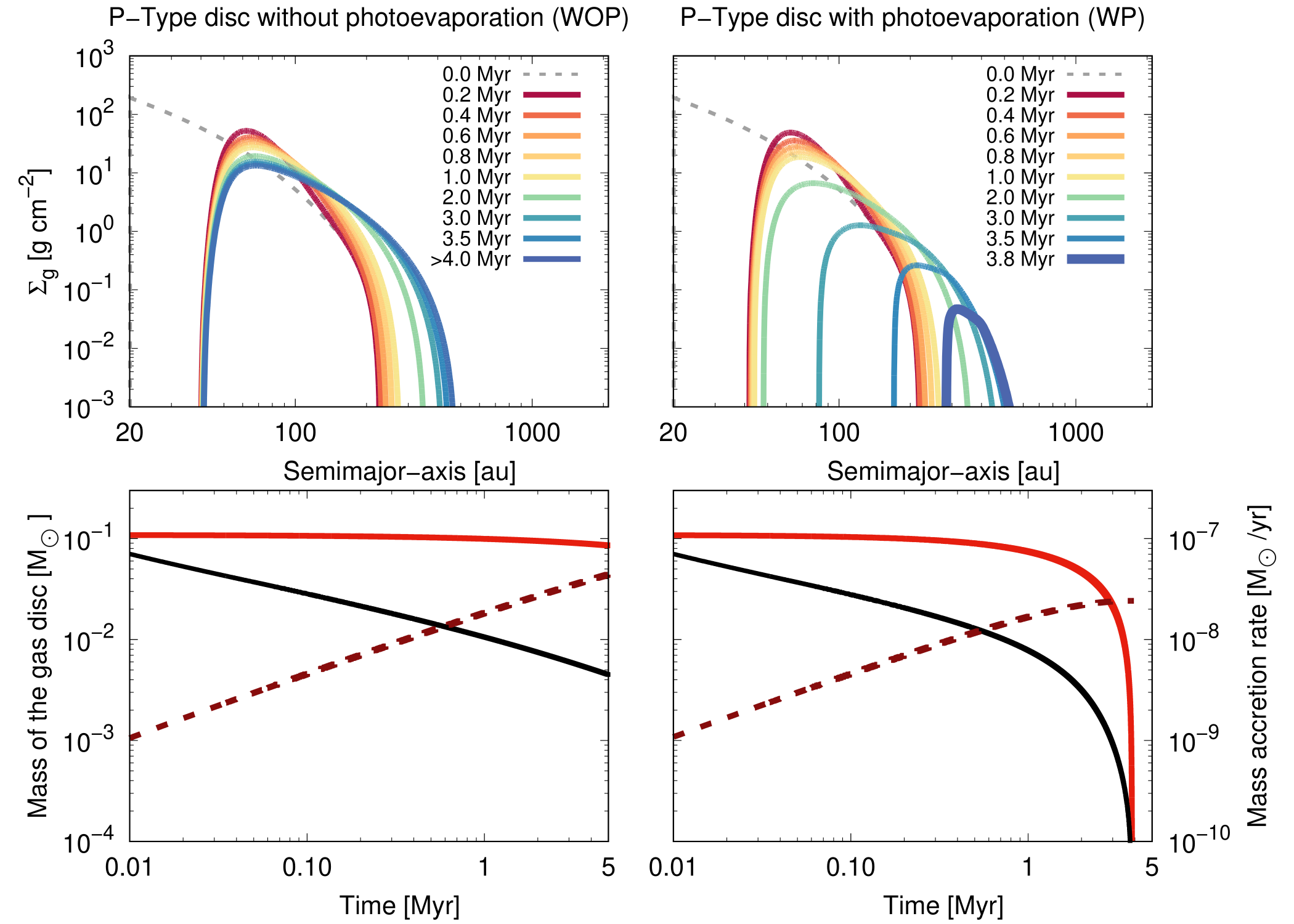}
    \caption{Time evolution of the CB gas disc. Left: without X-ray photoevaporation (WOP); right: including it (WP). Top panels show gas surface density profiles at different times; in the WP case, the thick blue line corresponds to the last profile before complete dissipation, and gray dashed lines indicate the initial profile. Bottom panels show the disc mass (solid red), cumulative mass lost by viscous accretion (dashed dark red), and mass accretion rate through tidal streams (solid black; range on the right-hand y-axis).}
  \label{Gas_Disco_P}
\end{figure*}

Modeling the evolution of the CB disc allows for a more realistic description of how material is transported toward the CP disc. A simpler approach would be to assume a constant mass injection rate, but this would miss the time-dependent nature of the process. Figure \ref{Gas_Disco_P} shows the time evolution of the CB gas surface density, $\Sigma^{\text{CB}}_{\text{g}}$, for the two modeled scenarios: without X-ray photoevaporation from the inner binary (left panel) and with photoevaporation (right panel). Hereafter, we refer to these cases as WOP and WP, respectively.

The main differences between these two extreme scenarios lie in the shape of the gas surface density profiles and in the disc dissipation timescale. From the beginning, in both the WOP and WP cases, the inner radius of the gas disc expands due to the angular momentum injected by the binary, pushing the disc outward and creating an inner cavity of up to $\sim$40 au (i.e., $\sim 2a_{\text{B}}$), in good agreement with \citet{ArtymowiczLubow1994}. It is important to note that the cavity size can vary significantly with the eccentricity and inclination of the binary system \citep{ArtymowiczLubow1994, MirandaLai2015, Miranda2017}. Moreover, if the disc is sufficiently massive for self-gravity to become important, the cavity may shrink toward the binary separation \citep{Mutter2017}. These effects, however, cannot be captured due to the one-dimensional nature of our model.

During the first million year, the evolution is nearly identical in both cases. However, from that point onward, when photoevaporation begins to be effective, the profiles become increasingly different. In the WOP case, the surface density profiles evolve slowly, with only minor changes in shape. The density gradually decreases as a result of the slow loss of mass through tidal streams, and the outer edge of the disc expands outward due to viscous evolution. In contrast, the disc affected by photoevaporation evolves faster: not only does the density decline faster and the outer edge expands but also the inner edge moves outward as a result of the efficient removal of material by photoevaporation, generating a growing cavity. As a consequence, the disc in the WP case completely dissipates within 3.8 Myr while the disc in the WOP case still has $\sim$82\% of its initial mass at the same time, meaning that it could provide gas and solids to the inner CP disc for a long time.

The difference between the mass evolution of both discs can also be appreciated in the bottom panels of Fig. \ref{Gas_Disco_P}. The gas disc mass is represented by solid red curves, whereas solid black curves show the mass accretion rate due to tidal streams and dashed dark red curves show the accumulated accreted mass in the WOP and WP cases. Note here that the registered mass accretion rate will be the corresponding mass injection rate into the CP disc. 

In the WOP case, we obtain accretion rates in the range 
$[2.25 \times 10^{-7},\, 4.5 \times 10^{-9}]\,M_\odot\,\mathrm{yr}^{-1}$,
with a mean value of $\sim 5.5 \times 10^{-9}\,M_\odot\, \mathrm{yr}^{-1}$. For the WP case, the accretion rate ranges between 
$[2.25 \times 10^{-7},\, 6.08 \times 10^{-11}]\,M_\odot\,\mathrm{yr}^{-1}$, with a mean value of $\sim 3.94 \times 10^{-9}\,M_\odot\,\mathrm{yr}^{-1}$. \citet{MarzariDangelo2025} reported an average gas accretion rate onto the CP disc of 
$\sim 2 \times 10^{-8}\,M_\odot\,\mathrm{yr}^{-1}$. 
Our mean accretion rates are therefore lower by a factor of $\sim 4$–5, although they remain within the same order of magnitude. This difference suggests that our model may underestimate the mass transfer, which could in principle affect the efficiency of planet formation. The implications of these accretion rates are explored in Sect. \ref{sec:in_situ}.

Finally, given the uncertainty as to whether the CB disc is indeed subject to photoevaporation, possibly mitigated by the shielding effect of the CP and CS discs, these two cases should be regarded as limiting scenarios, with the actual behavior likely lying somewhere in between. 

\subsubsection{The CP disc evolution: gaseous component}

\begin{figure*}[ht!]
  \centering
    \includegraphics[angle= 0, width= 0.9\textwidth]{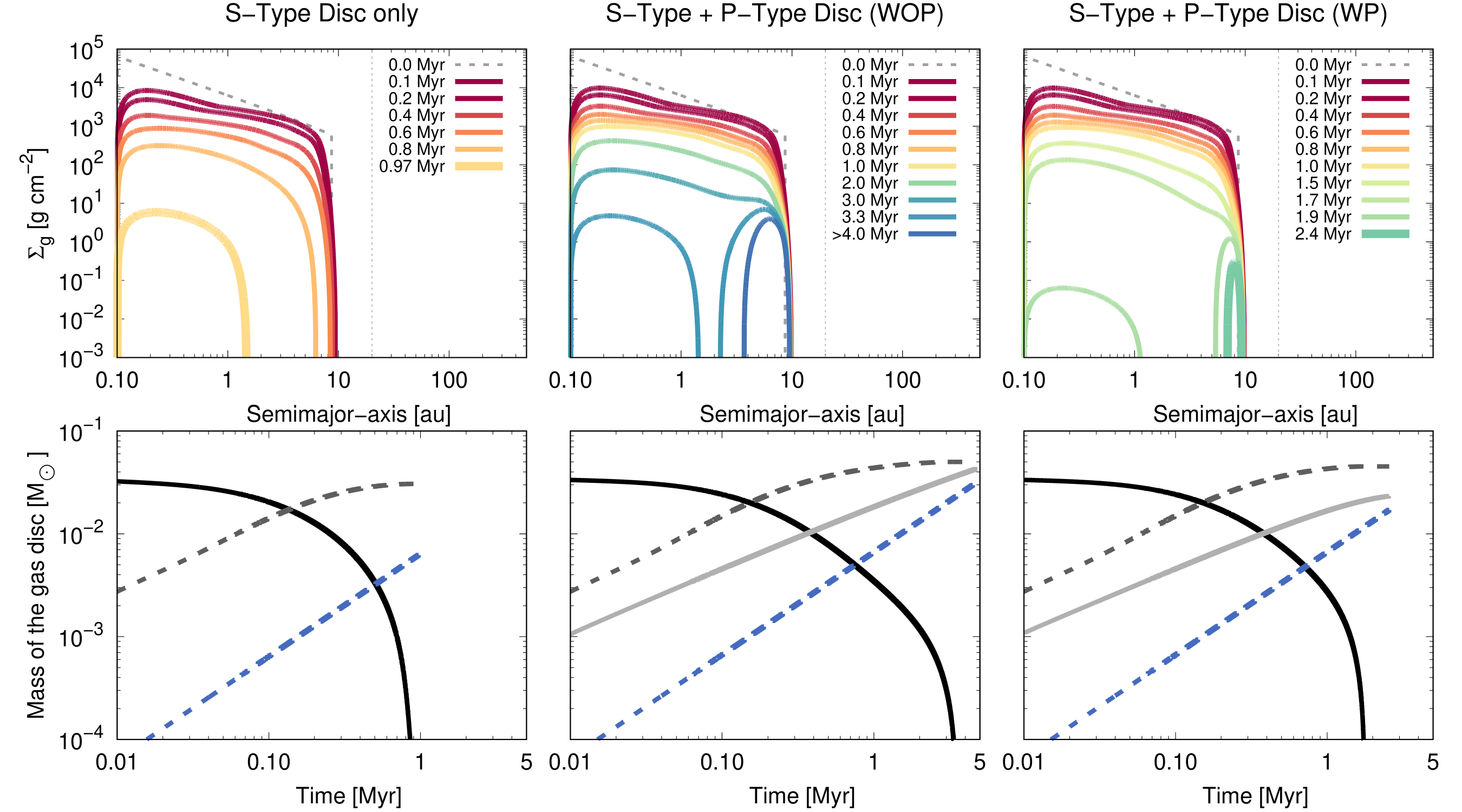}
    \caption{Top panels: time evolution of the gas surface density for the three cases: isolated CP disc (left), with gas injection from the CB disc without photoevaporation (WOP, middle), and with photoevaporation (WP, right). In the left and right panels, thick curves show the last profile before complete dissipation. Bottom panels: time evolution of the disc gas mass (black solid), cumulative mass lost by viscous accretion (gray dashed), cumulative mass lost by X-ray photoevaporation (blue dashed), and, for the WOP and WP cases, injected gas mass from the CB disc (light-gray solid).}
  \label{Gas_Disco_S}
\end{figure*}

Once the gas mass injection rate (which is equal to the CB mass accretion rate) from the CB disc onto the CP disc is computed, we can now model the evolution of the CP disc itself. Figure \ref{Gas_Disco_S} is analogous to Fig. \ref{Gas_Disco_P}, but for the CP gas surface density evolution, $\Sigma^{\text{CP}}_{\text{g}}$, and presents the results for three different configurations.
The left column shows the evolution of the CP disc in isolation. The middle column includes mass transfer from the CB disc without X-ray photoevaporation (WOP), while the right column shows the same configuration with photoevaporation included (WP). In all cases, X-ray photoevaporation from the primary star is considered.
 
The left column shows the fast dispersal (in less than 1 Myr) of the CP disc when it evolves in isolation. The disc lifetime and overall evolutionary pattern are broadly consistent with the results of \citet{RosottiClarke2018} (see the bottom panel of their Fig. 2), despite the differences in the adopted binary parameters. As they report, discs truncated at only a few astronomical units tend to disperse in an outside-in fashion, a behavior that is also evident in our simulations. This evolution is not only a consequence of the small disc size, but also of the strong X-ray photoevaporation, which efficiently removes the limited amount of gas located between the typical gap-opening radius and the outer edge of the disc.

A similar configuration was recently explored by \citet{Venturini2026} (see the top left panel of their Fig. 1). Although their results do not exhibit the same outside-in dispersal, this difference is likely related to the photoevaporation model adopted in their work. In particular, the use of an EUV-driven photoevaporation model \citep{Clarke2001}, which is less effective at removing mass from the outer disc than X-ray photoevaporation, promotes gap formation and the development of small transition-like discs rather than a global outside-in clearing.

This picture changes completely if we look at the middle and right columns of Fig. \ref{Gas_Disco_S}, which show the results of disc evolution with gas injection from the outer CB disc. Not only does the dissipation timescale increases ($>4$ times for the WOP, and $\sim$2.4 times for the WP cases) but it also changes the evolution of the disc, from an outside-in to an inside-out fashion. This is a direct consequence of both the new gas injected that replenishes the mass of gas in the outer region and photoevaporation which, because of that replenishment, manages to open a gap. 

In our simulations, this effect produces small, compact transition discs which, in the WOP case, can survive for several Myr due to the balance between gas injection, viscous evolution, and photoevaporation. However, their mass remains very low ($\sim$2 Neptune masses), making long-term survival unlikely in realistic environments. The mass and lifetime of these discs depend primarily on the accretion rate supplied by the CB disc. If this supply is more efficient, the resulting transition discs could be more massive and longer-lived.

In the WP case, the competition, after the gap opening, between mass loss from the CP disc driven by photoevaporation and mass injection from the CB disc is eventually won by the former, once the injection rate becomes too low. Consequently, the CP disc dissipates entirely. In other words, although the CB disc continues delivering gas for up to $\sim$3.5~Myr (see botton left panel of fig. \ref{Gas_Disco_P}), the supply is insufficient to maintain the small transition disc around the primary star, since photoevaporation efficiently removes all the incoming material.

The bottom panels of Fig.~\ref{Gas_Disco_S} show the evolution of the disc mass (solid black curves), the cumulative mass accreted onto the primary star (dark gray dashed lines), the cumulative mass lost by X-ray photoevaporation (blue dashed lines), and, for the WOP and WP cases, the cumulative gas mass injected from the CB disc. In the WOP case, the injected mass at $\sim4$ Myr amounts to $\sim1.1$ times the initial mass of the CP disc. In the WP case, by the time the disc is fully dispersed (at $\sim2.4$ Myr), the injected mass reaches $\sim0.85$ times the initial CP disc mass. This difference arises from the photoevaporation of the CB disc in the WP case, which becomes significant after $\sim1$ Myr. Interestingly, even in the WOP case, where the injected mass is comparable to the initial mass of the CP disc, the disc mass never increases. This indicates that viscous accretion together with X-ray photoevaporation remain the dominant processes controlling the disc evolution and lifetime.
 
\subsubsection{The CP disc evolution: solid component}

\begin{figure*}[ht!]
  \centering
    \includegraphics[angle= 0, width= 0.8\textwidth]{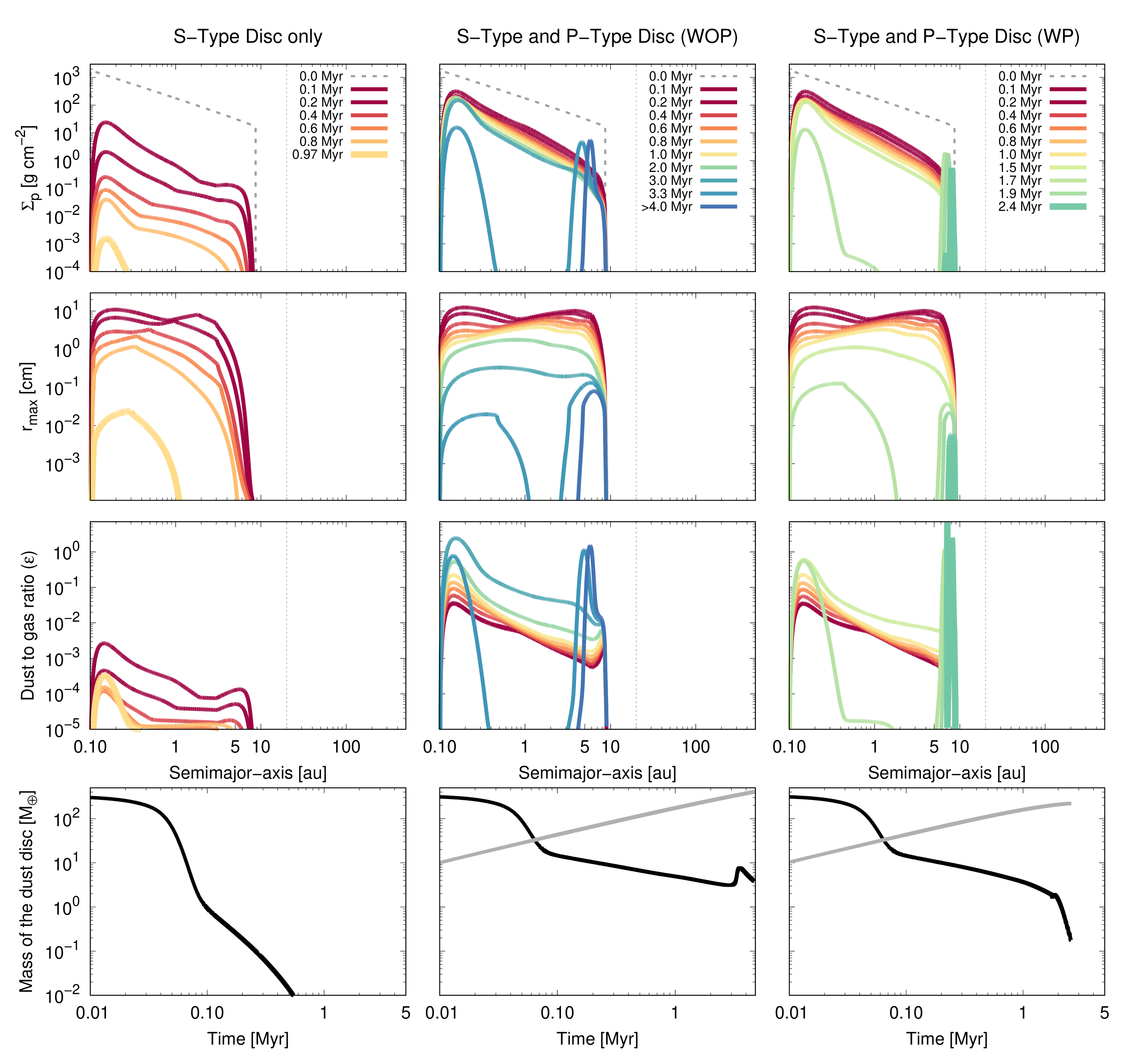}
    \caption{Results of the dust evolution for the isolated disc (left column) and the WOP and WP cases (middle and right columns). The first three rows show the evolution of the dust surface density, the maximum pebble size and the dust-to-gas ratio. The thick profiles show the last registered profile before complete gas disc dissipation. The bottom row shows, for each case, the time evolution of the mass of solids in black solid lines and the accumulated mass of solids injected coming from the CB disc in grey, only for the WOP and WP cases.}
  \label{fig:Polvo_Disco_S}
\end{figure*}

We now focus on describing the time evolution of the solid component of the CP disc in the three different simulated cases: the disc in isolation, the WOP and WP cases. The initial mass of solids for these cases is of $\sim320M_\oplus$.

Figure \ref{fig:Polvo_Disco_S} presents in the first three rows, from top to bottom, the evolution of the dust surface density $\Sigma^{\text{CP}}_{\text{d}}$, the maximum pebble size $r_{\text{max}}$, and the dust-to-gas ratio $\epsilon$. The last row shows the evolution of the total mass of dust/pebbles in solid black curves and the accumulated injected mass of dust for the WOP and WP cases in gray.  

In the isolated disc, we observe a rapid evolution driven by the extremely fast radial drift of solids. As also reported by \citet{Zagaria2021} and \citet{Venturini2026}, dust is removed much more efficiently than gas (and even more so than in circumstellar discs), falling by nearly two orders of magnitude in only $\sim$0.1 Myr. The depletion proceeds so quickly that photoevaporation, though relevant at later times, has no effect on the dust evolution in this scenario. This can also be appreciated by looking at the dust-to-gas ratio, which is lower than $10^{-4}$ within the first 0.2 Myr of evolution.  
Particles of initial size of 1 $\mu$m reach sizes of the order of 10~cm. However, as time advances, they decrease, and decrease faster in the outer disc regions. As in \citet{Zagaria2021}, the inner disc is fragmentation-limited while the outer regions are dominated by radial drift, although in our case drift clearly prevails overall. The separation between the two regimes is clearly marked by a change in the slope of the profiles. Similar to their findings, the maximum grain size limited by radial drift decrases faster than the one limited by fragmentation, which remains the dominant barrier in the inner disc. As a result, dust growth is more limited in the outer regions; nevertheless, the particles remain large enough to avoid being strongly coupled to the gas and thus are subject to significant radial drift. As a consequence of this evolution, the mass of solids decreases from $\sim320$~M$_{\oplus}$ to $\sim$1~M$_\oplus$ within the first 0.1 Myr, as can be appreciated in the left bottom panel of Fig. \ref{fig:Polvo_Disco_S}, severely affecting the chances of forming massive cores through pebble accretion (see section \ref{sec:in_situ}).
 
This situation, similar to what occurs for the gas, changes dramatically in the WOP and WP cases (middle and right panels of figure \ref{fig:Polvo_Disco_S}) where the CP disc is replenished with gas and dust from the external CB disc. In both scenarios, and in clear contrast to the isolated disc case, the dust surface density decreases by only $\sim$1-2 orders of magnitude within $\sim$0.1 Myr and, more importantly, remains roughly that level for about 1 Myr. This behavior is a direct consequence of the continuous injection of new solids that compensates for the rapid radial drift.  

During the first Myrs of evolution, and while the disc is complete, pebbles can reach the inner regions, increasing the pebble flux and potentially promoting the formation of massive cores (see Section \ref{sec:in_situ}). After about a few Myrs, when photoevaporation becomes significant, a gap also opens in the dust density as a consequence of the gap in the gas disc. The pebbles in the inner disc are rapidly lost mainly due to radial drift, while those in the outer region accumulate at the outer gas disc pressure maximum, forming a ring, and promoting a high dust-to-gas ratio at this location, which could contribute to planetesimal formation through streaming instability \citep{Youdin2005,Johansen2007}.

The continuous injection of dust into the disc also helps keeping the maximum particle size approximately uniform throughout the disc during at least the first 3 Myr in the WOP case and 2 Myr in the WP case, preventing a clear distinction between the fragmentation and radial-drift regimes.
This difference becomes evident again once the gap opens and the inner disc is no longer replenished with material, causing it to evolve rapidly as in the isolated case (see the inner light-blue profile at 3.3 Myr in the WOP case, comparable to the 0.97 Myr yellow profile in the isolated disc case). Meanwhile, the ongoing dust supply sustains the outer disc for a longer time. 

Finally, the mass of solids in the WOP and WP cases decrease at a clear lower peace than in the isolated disc case from $~320M_\oplus$ to $\sim3M_\oplus$ and to $1M_\oplus$ in $\sim$2 Myr, respectively (see middle and right bottom panels in fig. \ref{fig:Polvo_Disco_S}). The increase of solid material seen in the WOP case at approximately 3.5 Myr is due to the dust accumulation in the outer ring. In the WP case, this buildup is much less pronounced because the injection of dust is reduced as a consequence of the CB disc photoevaporation. The solid grey lines in these cases show the accumulated solid mass injected, that ends to be of the order of the initial mass of solids of the CP disc.

\subsubsection{In-situ planet formation}
\label{sec:in_situ}

In this section, we describe the in situ growth of a single moon-mass embryo through pebble and gas accretion at fixed locations within the three previously modeled scenarios. The main goal here is to assess whether the injection of gas and dust into the CP disc, supplied by the external CB disc, can facilitate the formation of more massive planets in strongly truncated discs of close binaries, thereby potentially overcoming the challenges of planet formation in such environments.

\begin{figure*}[ht!]
  \centering
    \includegraphics[angle=0, width= 0.9\textwidth]{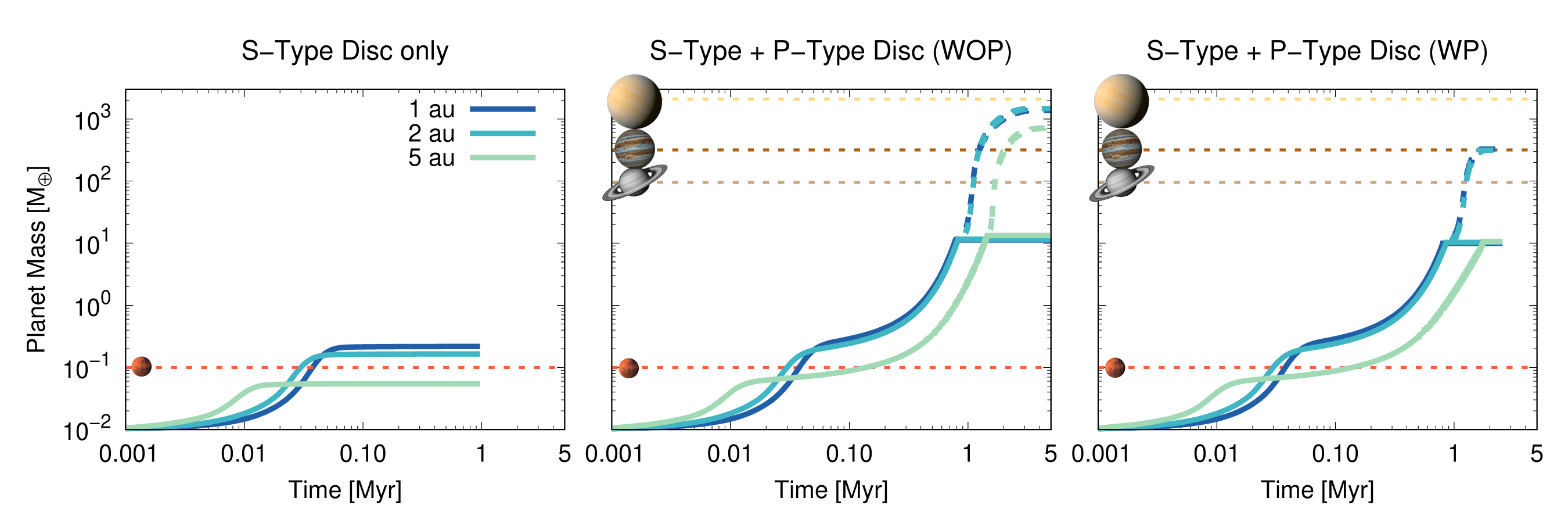}
    \caption{In-situ planet growth by pebble and gas accretion at 1, 2, and 5 au in the CP disc. Each curve corresponds to an independent planet-formation run (one embryo per run). Thick solid lines show core growth, and thick dashed lines the total mass (core + envelope). The masses of Mars, Saturn, Jupiter, and $\gamma$-Cephei Ab \citep{Knudstrup2023} are indicated by thin dashed lines.}
  \label{Formacion_Planetaria}
\end{figure*}

Figure \ref{Formacion_Planetaria} presents the time evolution of the mass of the core in solid lines, and the total mass of the planet (core and envelope) in dashed lines. Each planet, modeled individually and in isolation, grows in situ at the selected locations of 1, 2, and 5 au, spanning the extent of the CP disc. The masses of Mars, Saturn, Jupiter and $\gamma$-Cephei Ab are marked for reference.

The left panel shows planets growing in the isolated disc. The innermost planets, at 1 and 2 au, grow only up to between 2-3 Mars masses ($M_{\text{Mars}}\sim0.1M_\oplus$), and they accrete solids only during the first $\sim$50.000 years. Beyond this point, the solid reservoir declines significantly. The outer planet, located at 5 au, grows even less because the local dust surface density is lower and it only receives the pebble flux from beyond its orbit. By contrast, the planets at 1 and 2 au achieve slightly larger growth, as they intercept a greater pebble flux. These results are in agreement with those presented in \citet{Venturini2026} (see the light-blue solid line in the top panel of their fig. 4).

We now focus on the middle and right panels of Fig. \ref{Formacion_Planetaria}, which present the WOP and WP cases. In both scenarios, during the first $\sim$50.000 years, the three planets evolve similarly to those in the isolated case. Afterwards, however, the sustained pebble flux allows them to grow significantly more.

In the WOP case, all three cores reach their pebble isolation mass and stop accreting pebbles, doing so earlier at smaller orbital distances since this mass increases with distance from the star. As a consequence, gas accretion is triggered while a substantial amount of gas is still available in the disc, leading to the formation of gas giants with final masses between Jupiter and the most updated mass of $\gamma$-Cephei Ab, 6.6$M_{\text{Jup}}$ \citep{Knudstrup2023}. 

In the WP case, although the innermost planets reach their pebble isolation mass, they do so later than in the previous case. They are still able to accrete large amounts of gas but do not grow beyond a Jupiter-mass planet. By contrast, the outermost planet never reaches its pebble isolation mass and remains a Super-Earth of $\sim10,M_\oplus$.

As noted previously, and given the current uncertainty regarding whether the potential CB disc could undergo photoevaporation, the actual conditions for planet formation likely lie somewhere between these two scenarios. 

We also note that, despite the lower accretion rates from the CB disc onto the CP disc obtained in our model compared to those reported by \citet{MarzariDangelo2025} (as discussed in sec. \ref{sec:CBdiscevolution}), we are still able to form giant planets. However, it is important to emphasize that these results may depend on the initial conditions adopted, such as the fragmentation velocity threshold for dust growth and the initial mass of the system, which determine the amount of gas and solids available for planet formation. We explore different values for these parameters in Appendix \ref{App:Vfrag} and \ref{App:Msyst}.


\subsection{The eccentric case}
\label{sec:ecc}

We now present the results for the more realistic eccentric configuration, where the eccentricity of $\gamma$~Cephei is included in both the CP disc truncation and the evolution of the CB disc.

\begin{figure*}[ht]
  \centering
    \includegraphics[angle= 0, width= 0.9\textwidth]{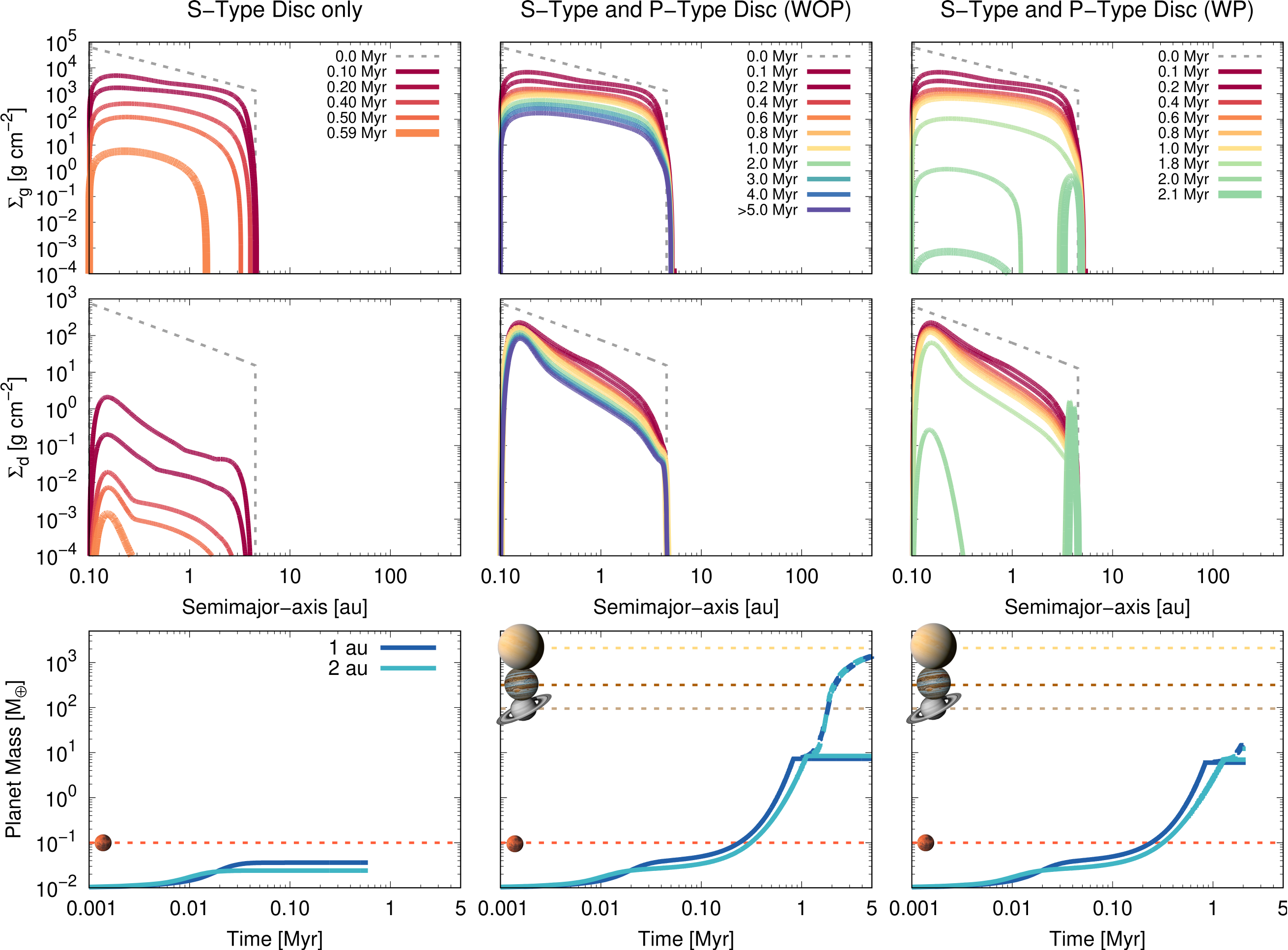}
    \caption{Time evolution of the gas surface density (top row) and dust surface density profiles (mid row) for the isolated disc (left), and the WOP and WP cases (middle and right). The bottom row shows the planet formation tracks for planets growing in-situ at 1 and 2 au for each case. As in figure \ref{Formacion_Planetaria}, solid lines show the core growth and the dashed lines the total mass of the planet (core and envelope). Also here, the masses of Mars, Saturn, Jupiter and $\gamma$-Cephei b \citep{Knudstrup2023} are marked with thin dashed lines for reference.}
  \label{fig:Formacion_Planetaria_exc}
\end{figure*}

Figure \ref{fig:Formacion_Planetaria_exc} shows the time evolution of the gas and dust surface density profiles (top and middle rows, respectively) and the corresponding planet formation tracks (bottom rows) for the isolated disc, and for the WOP and WP cases.

The main difference from the circular fiducial case is that the stronger truncation caused by the eccentric binary reduces the initial gas mass and dust in the CP disc by about $\sim$46\% (see Table \ref{tab:1}). The initial dust mass for this case is of $\sim177M_\oplus$. As a direct consequence of this mass deficit, the isolated disc (left column) evolves extremely fast. The gaseous component dissipates in less than $\sim0.6$ Myr, while the dust surface density decreases by nearly three orders of magnitude in less than $0.1$ Myr. This leads to an even less efficient planet formation process than in the circular case (see bottom left panel). Our planets do not even reach the mass of Mars. We note that we only compute the formation of planets in-situ at 1 and 2 au since the CP disc truncates at 4.53 au.

One might expect that, in the eccentric WOP and WP cases, the disc lifetimes would lie between those of the eccentric isolated case and their circular counterparts. However, this is not what we find, at least for the WOP case. Instead, these discs show longer-than-expected lifetimes, and the planets that form reach masses comparable to those obtained in the circular case, at least in the WOP case.

This apparently counterintuitive result is a direct consequence of X-ray photoevaporation from the central star, which is effective only beyond the gravitational radius. When the CP disc is truncated close to this location, the amount of material removed by photoevaporation is significantly reduced. The disc evolution is then dominated by viscous accretion, which effectively extends its lifetime. In particular, in the WOP case, the mass removed by photoevaporation is not sufficient to open a gap in the disc, at least within the 5 Myr covered by our simulations, during which the CP disc continuously receives gas and dust from the CB disc. As a consequence, planets forming in situ at 1 and 2 au are able to accrete more material: the gap never opens to halt the pebble flux toward the inner regions, and the sustained supply of solids allows them to grow up to a few Jupiter masses. 

In the WP case and as a consequence of a reduction in the mass injection due to the CB disc photoevaporation, the disc manages to open a gap and the pebble flux towards the inner planets stops. Despite in this case both planets reach their pebble isolation masses, they do it too late, almost while there is no more gas to accrete, and they end as Mini-Neptunes. Nevertheless, the important result is that despite the extreme disc truncation, this mechanism can still form giant planets.

As a summary of our results, table \ref{tab:2} provides the planet formation outcomes, both for the circular and eccentric case, for each simulated scenario. 

\begin{table*}
\centering
\renewcommand{\arraystretch}{1.5}
\begin{tabular}{lccc} 
\hline 
 & S-type Disc only & S-type + P-type Disc (WOP) & S-type + P-type Disc (WP) \\
\hline
Circular case & 0.05 - 0.21  & 725 - 1476  &  10.6 - 328  \\
\hline
Eccentric case & 0.024 - 0.036 & 1611 - 1640 & 12.2 - 14.6 \\
\hline
\end{tabular}
    \vspace{0.2cm}
    \caption{Ranges of total mass (core and envelope) of the planets (in  M$_\oplus$) formed in our simulations for the circular and eccentric cases. Note that the most updated mass of $\gamma$-Cephei Ab is $\sim6.6M_{\text{Jup}}$ \citep{Knudstrup2023}, and our most massive simulated planet reaches $5.15M_{\text{Jup}}$.} 
    \label{tab:2}
\end{table*}

\section{Discussion}
\label{sec:discussion}

In this work we explore the consequences of a potentially efficient mechanism of mass transfer from an external CB disc to a truncated CP disc in a $\gamma$-Cephei–like system, and how this process may alter the outcome of planet formation by pebble and gas accretion. While our results show that this mechanism can dramatically change the ability of truncated discs to form massive planets, both the modelling approach and the physical scenario itself deserve careful discussion.

\subsection{Model caveats}

Our simulations are based on a 1D+1D viscous evolution model for both the CP and CB discs \citep{Ronco2021}, coupled with a planet formation module via pebble and gas accretion \citep{Guilera2020,Venturini2020SE,Venturini2020Letter}. Although this approach does not capture the intrinsically multidimensional structure expected in discs within close binary systems—where gas streams, azimuthal asymmetries, and other non-axisymmetric features may arise \citep{MarzariDangelo2025}—it enables us to follow the long-term coupled evolution of both discs and to model planet formation within this evolving environment. Fully hydrodynamical simulations including both disc dynamics and planet formation over Myr timescales remain prohibitively expensive, particularly when exploring different system parameters as done here.

In this work, the hydrodynamical transport of material across the CB cavity is represented through a mass injection rate derived from the CB disc evolution. This evolution depends on the adopted initial conditions, particularly the total mass in the system and the initial gas surface density profile, which determine the efficiency and duration of the material supply to the CP disc and thus the planet formation outcomes. Different initial masses or surface density distributions would therefore lead to different evolutionary pathways. However, our aim is not to perform an exhaustive parameter study, but to explore whether this replenishment mechanism can enable giant planet formation under $\gamma$ Cephei–like conditions.

We note that we did not compute the dust growth and evolution on the CB disc, and assume a constant dust-to-gas ratio. This simplification allows us to link the solid mass supply directly to the gas flow, focusing on how the global gas disc evolution governs the delivery of material to the CP disc. However, this might not be correct since, in general, the dust-to-gas ratio do not remain constant on time during disc evolution. 
In addition, \citet{MarzariDangelo2025} found that dust filtration via streams deppends on the dust size also. They found that mm-size particles could be effeciently trapped at the CB disc cavity. This could also influence the dust growth and evolution in the CP disc, and hence on the pebble accretion rates.
 
As in \citet{Venturini2026}, our model does not include the direct gravitational interaction between the forming planets and the stellar companion. However, such interactions could excite planetary eccentricities, modify migration, and influence pebble accretion efficiencies, as recently shown by \citet{Nigioni2026}. By neglecting these effects, we intentionally isolate the role of the evolving disc structure, and in particular the external mass supply, regulating planetary growth. Nevertheless, it is intriguing to assess whether the eccentricity that a $\gamma-$Cephei-like planet may acquire in these systems is affected by a formation scenario such as the WOP case. We explore this aspect in Appendix \ref{App:Nbody}, and show that the planet's eccentricity evolves toward the forced value imposed by the binary, consistent with theoretical expectations.

We also do not include the effect of external photoevaporation from potentially nearby massive stars (\citealp[see][and references there in]{WinterHaworth2022}). This omission does not arise from a modelling limitation, but from the fact that the birth environment of systems such as $\gamma$ Cephei is not well known and may vary significantly from one system to another. 

Another deliberate choice of our framework is to assume that planets form in situ at 1, 2, and 5 au. In real binary systems, the locations of planetesimal formation, early core growth, and subsequent migration may be strongly influenced by the stellar companion and the corresponding planet-disc interactions \citep{Nigioni2026}. By adopting an in situ scenario, we do not attempt to reproduce the full planet formation pathway in binaries, but rather to explore how the evolving disc conditions affect planetary growth once planetary seeds are present.

Finally, X-ray photoevaporation rates  adopted in this work \citep{Owen2012} might be overestimated by at least an order of magnitude, as recently suggested by \citet{Sellek2024}. Lower photoevaporation rates would tend to prolong the lifetime of both the CB and CP discs. However, it is the lifetime of the dust component of the disc, rather than that of the gas, that is most relevant for planet formation \citep{Zagaria2021,Venturini2026}. In truncated CP discs, pebbles are rapidly depleted due to radial drift and the lack of an external reservoir, limiting core growth. Therefore, sustained dust/pebble replenishment is more critical than an extended gas disc lifetime in our scenario.

\subsection{Caveats of the proposed scenario}

Beyond our modelling choices, several uncertainties are associated with the scenario itself. One of the main ones concerns the existence and frequency of external CB discs in $\gamma$-Cephei–like systems. Although only a few examples have been reported so far, some systems show evidence of extended CB reservoirs, such as GG Tau A \citep{Dutrey1994,Dutrey2014} and L1551 IRS5 \citep{Maureira2026,Cuello2026}.

Another uncertainty concerns the efficiency of X-ray photoevaporation from the inner binary in dispersing the outer CB disc, as circumstellar discs around each star may partially shield this effect. In Appendix \ref{App:Disco_CS}, we estimate the dispersal timescale of a potential CS disc and argue that its rapid dissipation could reduce this shielding, placing the actual planet formation outcomes between the WOP and WP cases.

Likewise, the existence, mass, and lifetime of a CB disc capable of sustaining prolonged mass transfer to the CP and/or CS discs are likely to be system dependent. This is, its effectiveness will depend on parameters such as the binary mass ratio, separation, and binary eccentricity, which control both the truncation of the CP and CS discs, the structure of the CB disc and the potential mass transfer to the primary and secondary stars. Exploring a wider parameter space with dedicated hydrodynamical simulations would help to determine in which types of systems this process may, or may not, play a significant role.

For all these reasons, our results should therefore be interpreted as explorative. However, they clearly show that whenever sustained mass transfer from a CB disc to a truncated CP disc is possible, the prospects for forming massive planets by pebble and gas accretion change qualitatively. 
  
\section{Conclusions}

The aim of this work was to test whether the injection of gas and solids from a CB disc, remnant of the star formation process, can facilitate the formation of giant planets in the CP disc of $\gamma$-Cephei–like systems. To this end, we coupled the long-term evolution of the CP and CB discs with our planet formation model PLANETALP including pebble and gas accretion.

Our approach differs from previous hydrodynamical studies, such as \citet{MarzariDangelo2025}, which focused on the gas and dust dynamics of replenished CP discs but did not explicitly model the planet formation process. By including pebble accretion and the subsequent gas accretion phase, our simulations allow us to directly assess whether the extended disc lifetimes and additional solid reservoirs produced by CB mass transfer can lead to the formation of giant planets.

Our results show that sustained mass transfer from the CB disc can significantly modify the planet formation outcome. For the adopted parameters planet growth is strongly inhibited in isolated truncated discs and embryos fail to grow beyond a few Mars-mass planets. When gas and dust are injected from the CB disc, however, the CP disc can remain massive for longer timescales, even if affected by strong X-ray photoevaporation, allowing the formation of giant planets through pebble accretion followed by gas accretion. In particular, planets with masses comparable to $\gamma$-Cephei b can form in situ under favorable conditions, especially in those cases in which the potential CB disc photoevaporation is not efficient (WOP cases).

These results suggest that disc replenishment from an external CB reservoir may provide a viable pathway to overcome the strong truncation imposed by the stellar companion in close binary systems. Nevertheless, the efficiency of this mechanism depends on several uncertain factors, including the existence and lifetime of CB reservoirs and the efficiency of disc dispersal processes. A broader exploration of binary parameters and disc properties will therefore be required to determine how common this formation pathway may be.

\begin{acknowledgements}
 We thank the referee for their constructive comments and suggestions, which helped improve the clarity of the manuscript and strengthen the discussion. This work is partially supported by PIP 2971 from CONICET (Argentina). We also thank Juan Ignacio Rodriguez
from IALP for the computation managing resources of the Grupo de Astrofísica
Planetaria de La Plata.

\end{acknowledgements}

\bibliographystyle{aa}
\bibliography{S-and-P}

\begin{thebibliography}{88}
\expandafter\ifx\csname natexlab\endcsname\relax\def\natexlab#1{#1}\fi

\bibitem[{{Alexander}(2012)}]{Alexander2012}
{Alexander}, R. 2012, \apjl, 757, L29

\bibitem[{{Andrews} {et~al.}(2009){Andrews}, {Wilner}, {Hughes}, {Qi}, \&
  {Dullemond}}]{Andrews2009}
{Andrews}, S.~M., {Wilner}, D.~J., {Hughes}, A.~M., {Qi}, C., \& {Dullemond},
  C.~P. 2009, \apj, 700, 1502

\bibitem[{{Armitage} \& {Natarajan}(2002)}]{ArmitageNatarajan2002}
{Armitage}, P.~J. \& {Natarajan}, P. 2002, \apjl, 567, L9

\bibitem[{{Artymowicz} \& {Lubow}(1994)}]{ArtymowiczLubow1994}
{Artymowicz}, P. \& {Lubow}, S.~H. 1994, \apj, 421, 651

\bibitem[{{Artymowicz} \& {Lubow}(1996)}]{ArtymowiczLubow1996}
{Artymowicz}, P. \& {Lubow}, S.~H. 1996, \apjl, 467, L77

\bibitem[{{Baraffe} {et~al.}(2015){Baraffe}, {Homeier}, {Allard}, \&
  {Chabrier}}]{Baraffe2015}
{Baraffe}, I., {Homeier}, D., {Allard}, F., \& {Chabrier}, G. 2015, \aap, 577,
  A42

\bibitem[{{Beaug{\'e}} {et~al.}(2010){Beaug{\'e}}, {Leiva}, {Haghighipour}, \&
  {Otto}}]{Beauge2010}
{Beaug{\'e}}, C., {Leiva}, A.~M., {Haghighipour}, N., \& {Otto}, J.~C. 2010,
  \mnras, 408, 503

\bibitem[{{Beaug{\'e}} {et~al.}(2006){Beaug{\'e}}, {Michtchenko}, \&
  {Ferraz-Mello}}]{Beauge2006}
{Beaug{\'e}}, C., {Michtchenko}, T.~A., \& {Ferraz-Mello}, S. 2006, \mnras,
  365, 1160

\bibitem[{{Behmard} {et~al.}(2022){Behmard}, {Dai}, \& {Howard}}]{Behmard2022}
{Behmard}, A., {Dai}, F., \& {Howard}, A.~W. 2022, \aj, 163, 160

\bibitem[{{Birnstiel} {et~al.}(2012){Birnstiel}, {Klahr}, \&
  {Ercolano}}]{Birnstiel2012}
{Birnstiel}, T., {Klahr}, H., \& {Ercolano}, B. 2012, \aap, 539, A148

\bibitem[{{Birnstiel} {et~al.}(2011){Birnstiel}, {Ormel}, \&
  {Dullemond}}]{Birnstiel2011}
{Birnstiel}, T., {Ormel}, C.~W., \& {Dullemond}, C.~P. 2011, \aap, 525, A11

\bibitem[{{Brasser} {et~al.}(2017){Brasser}, {Bitsch}, \&
  {Matsumura}}]{Brasser2017}
{Brasser}, R., {Bitsch}, B., \& {Matsumura}, S. 2017, \aj, 153, 222

\bibitem[{{Camargo} {et~al.}(2023){Camargo}, {Kley}, \& {Winter}}]{Camargo2023}
{Camargo}, B.~C.~B., {Kley}, W., \& {Winter}, O.~C. 2023, \mnras, 522, 6394

\bibitem[{{Camargo} {et~al.}(2024){Camargo}, {Moraes}, {Winter}, \&
  {Foryta}}]{Camargo2024}
{Camargo}, B.~C.~B., {Moraes}, R.~A., {Winter}, O.~C., \& {Foryta}, D.~W. 2024,
  \mnras, 535, 3020

\bibitem[{{Clarke} {et~al.}(2001){Clarke}, {Gendrin}, \&
  {Sotomayor}}]{Clarke2001}
{Clarke}, C.~J., {Gendrin}, A., \& {Sotomayor}, M. 2001, \mnras, 328, 485

\bibitem[{{Cuadra} {et~al.}(2009){Cuadra}, {Armitage}, {Alexander}, \&
  {Begelman}}]{Cuadra2009}
{Cuadra}, J., {Armitage}, P.~J., {Alexander}, R.~D., \& {Begelman}, M.~C. 2009,
  \mnras, 393, 1423

\bibitem[{{Cuello} {et~al.}(2026){Cuello}, {Bianchi}, {M{\'e}nard}, {Loinard},
  {Hern{\'a}ndez Garnica}, {Dur{\'a}n}, {Ceccarelli}, {Maureira}, {Chandler},
  {Codella}, {Sakai}, {Podio}, {Sabatini}, {Chahine}, {de Simone}, {Fedele},
  {Johnstone}, {Hanawa}, {Jim{\'e}nez-Serra}, \& {Yamamoto}}]{Cuello2026}
{Cuello}, N., {Bianchi}, E., {M{\'e}nard}, F., {et~al.} 2026, \aap, 705, L16

\bibitem[{{Dr{\c a}zkowska} {et~al.}(2016){Dr{\c a}zkowska}, {Alibert}, \&
  {Moore}}]{Drazkowska2016}
{Dr{\c a}zkowska}, J., {Alibert}, Y., \& {Moore}, B. 2016, \aap, 594, A105

\bibitem[{{Dr{\k{a}}{\.z}kowska} \& {Alibert}(2017)}]{Drazkowska2017}
{Dr{\k{a}}{\.z}kowska}, J. \& {Alibert}, Y. 2017, \aap, 608, A92

\bibitem[{{Dunhill} {et~al.}(2015){Dunhill}, {Cuadra}, \&
  {Dougados}}]{Dunhill2015}
{Dunhill}, A.~C., {Cuadra}, J., \& {Dougados}, C. 2015, \mnras, 448, 3545

\bibitem[{{Dutrey} {et~al.}(2014){Dutrey}, {di Folco}, {Guilloteau}, {Boehler},
  {Bary}, {Beck}, {Beust}, {Chapillon}, {Gueth}, {Hur{\'e}}, {Pierens},
  {Pi{\'e}tu}, {Simon}, \& {Tang}}]{Dutrey2014}
{Dutrey}, A., {di Folco}, E., {Guilloteau}, S., {et~al.} 2014, \nat, 514, 600

\bibitem[{{Dutrey} {et~al.}(1994){Dutrey}, {Guilloteau}, \&
  {Simon}}]{Dutrey1994}
{Dutrey}, A., {Guilloteau}, S., \& {Simon}, M. 1994, \aap, 286, 149

\bibitem[{{Endl} {et~al.}(2011){Endl}, {Cochran}, {Hatzes}, \&
  {Wittenmyer}}]{Endl2011}
{Endl}, M., {Cochran}, W.~D., {Hatzes}, A.~P., \& {Wittenmyer}, R.~A. 2011, in
  American Institute of Physics Conference Series, Vol. 1331, Planetary Systems
  Beyond the Main Sequence, ed. S.~{Schuh}, H.~{Drechsel}, \& U.~{Heber} (AIP),
  88--94

\bibitem[{{Farris} {et~al.}(2014){Farris}, {Duffell}, {MacFadyen}, \&
  {Haiman}}]{Farris2014}
{Farris}, B.~D., {Duffell}, P., {MacFadyen}, A.~I., \& {Haiman}, Z. 2014, \apj,
  783, 134

\bibitem[{{Fontecilla} {et~al.}(2019){Fontecilla}, {Haiman}, \&
  {Cuadra}}]{Fontecilla2019}
{Fontecilla}, C., {Haiman}, Z., \& {Cuadra}, J. 2019, \mnras, 482, 4383

\bibitem[{{Giuppone} {et~al.}(2011){Giuppone}, {Leiva}, {Correa-Otto}, \&
  {Beaug{\'e}}}]{Giuppone2011}
{Giuppone}, C.~A., {Leiva}, A.~M., {Correa-Otto}, J., \& {Beaug{\'e}}, C. 2011,
  \aap, 530, A103

\bibitem[{{Guilera} {et~al.}(2017){Guilera}, {Miller Bertolami}, \&
  {Ronco}}]{Guilera2017}
{Guilera}, O.~M., {Miller Bertolami}, M.~M., \& {Ronco}, M.~P. 2017, \mnras,
  471, L16

\bibitem[{{Guilera} {et~al.}(2020){Guilera}, {S{\'a}ndor}, {Ronco},
  {Venturini}, \& {Miller Bertolami}}]{Guilera2020}
{Guilera}, O.~M., {S{\'a}ndor}, Z., {Ronco}, M.~P., {Venturini}, J., \& {Miller
  Bertolami}, M.~M. 2020, arXiv e-prints, arXiv:2005.10868

\bibitem[{{Gundlach} \& {Blum}(2015)}]{GundlachBlum2015}
{Gundlach}, B. \& {Blum}, J. 2015, \apj, 798, 34

\bibitem[{{Hatzes} {et~al.}(2003){Hatzes}, {Cochran}, {Endl}, {McArthur},
  {Paulson}, {Walker}, {Campbell}, \& {Yang}}]{Hatzes2003}
{Hatzes}, A.~P., {Cochran}, W.~D., {Endl}, M., {et~al.} 2003, \apj, 599, 1383

\bibitem[{{Heppenheimer}(1978)}]{Heppenheimer1978}
{Heppenheimer}, T.~A. 1978, \aap, 65, 421

\bibitem[{{Huang} \& {Ji}(2022)}]{HuangJi2022}
{Huang}, X. \& {Ji}, J. 2022, \aj, 164, 177

\bibitem[{{Ikoma} {et~al.}(2000){Ikoma}, {Nakazawa}, \& {Emori}}]{Ikoma2000}
{Ikoma}, M., {Nakazawa}, K., \& {Emori}, H. 2000, \apj, 537, 1013

\bibitem[{{Johansen} {et~al.}(2007){Johansen}, {Oishi}, {Mac Low}, {Klahr},
  {Henning}, \& {Youdin}}]{Johansen2007}
{Johansen}, A., {Oishi}, J.~S., {Mac Low}, M.-M., {et~al.} 2007, \nat, 448,
  1022

\bibitem[{{Knudstrup} {et~al.}(2023){Knudstrup}, {Lund}, {Fredslund Andersen},
  {R{\o}rsted}, {P{\'e}rez Hern{\'a}ndez}, {Grundahl}, {Pall{\'e}}, {Stello},
  {White}, {Kjeldsen}, {Vrard}, {Winther}, {Handberg}, \&
  {Sim{\'o}n-D{\'\i}az}}]{Knudstrup2023}
{Knudstrup}, E., {Lund}, M.~N., {Fredslund Andersen}, M., {et~al.} 2023, \aap,
  675, A197

\bibitem[{{Lambrechts} \& {Johansen}(2012)}]{Lambrechts2012}
{Lambrechts}, M. \& {Johansen}, A. 2012, \aap, 544, A32

\bibitem[{{Lambrechts} \& {Johansen}(2014)}]{LambrechtsJohansen2014}
{Lambrechts}, M. \& {Johansen}, A. 2014, \aap, 572, A107

\bibitem[{{Lambrechts} {et~al.}(2014){Lambrechts}, {Johansen}, \&
  {Morbidelli}}]{Lambrechts+2014}
{Lambrechts}, M., {Johansen}, A., \& {Morbidelli}, A. 2014, \aap, 572, A35

\bibitem[{{Lester} {et~al.}(2021){Lester}, {Matson}, {Howell}, {Furlan},
  {Gnilka}, {Scott}, {Ciardi}, {Everett}, {Hartman}, \& {Hirsch}}]{Lester2021}
{Lester}, K.~V., {Matson}, R.~A., {Howell}, S.~B., {et~al.} 2021, \aj, 162, 75

\bibitem[{{Liu} \& {Ormel}(2018)}]{Liu2018}
{Liu}, B. \& {Ormel}, C.~W. 2018, \aap, 615, A138

\bibitem[{{Lodders} {et~al.}(2025){Lodders}, {Bergemann}, \&
  {Palme}}]{Lodders2025}
{Lodders}, K., {Bergemann}, M., \& {Palme}, H. 2025, \ssr, 221, 23

\bibitem[{{Mart{\'\i}} \& {Beaug{\'e}}(2012)}]{MartiBeauge2012}
{Mart{\'\i}}, J.~G. \& {Beaug{\'e}}, C. 2012, \aap, 544, A97

\bibitem[{{Marzari} \& {D'Angelo}(2025)}]{MarzariDangelo2025}
{Marzari}, F. \& {D'Angelo}, G. 2025, \aap, 695, A53

\bibitem[{{Marzari} \& {Scholl}(2000)}]{MarzariScholl2000}
{Marzari}, F. \& {Scholl}, H. 2000, \apj, 543, 328

\bibitem[{{Marzari} \& {Thebault}(2019)}]{MarzariThebault2019}
{Marzari}, F. \& {Thebault}, P. 2019, Galaxies, 7, 84

\bibitem[{{Maureira} {et~al.}(2026){Maureira}, {Pineda}, {Liu}, {Caselli},
  {Chandler}, {Testi}, {Johnstone}, {Segura-Cox}, {Loinard}, {Bianchi},
  {Codella}, {Miotello}, {Podio}, {Cacciapuoti}, {Oya}, {Lopez-Sepulcre},
  {Sakai}, {Zhang}, {Cuello}, {Ohashi}, {Aikawa}, {Sabatini}, {Zhang},
  {Ceccarelli}, \& {Yamamoto}}]{Maureira2026}
{Maureira}, M.~J., {Pineda}, J.~E., {Liu}, H.~B., {et~al.} 2026, \aap, 705, A96

\bibitem[{{Miranda} \& {Lai}(2015)}]{MirandaLai2015}
{Miranda}, R. \& {Lai}, D. 2015, \mnras, 452, 2396

\bibitem[{{Miranda} {et~al.}(2017){Miranda}, {Mu{\~n}oz}, \&
  {Lai}}]{Miranda2017}
{Miranda}, R., {Mu{\~n}oz}, D.~J., \& {Lai}, D. 2017, \mnras, 466, 1170

\bibitem[{{Mugrauer} \& {Michel}(2021)}]{Mugrauer2021}
{Mugrauer}, M. \& {Michel}, K.-U. 2021, Astronomische Nachrichten, 342, 840

\bibitem[{{Mugrauer} {et~al.}(2023){Mugrauer}, {R{\"u}ck}, \&
  {Michel}}]{Mugrauer2023}
{Mugrauer}, M., {R{\"u}ck}, J., \& {Michel}, K.-U. 2023, Astronomische
  Nachrichten, 344, e20230055

\bibitem[{{M{\"u}ller} \& {Kley}(2012)}]{MullerKley2012}
{M{\"u}ller}, T.~W.~A. \& {Kley}, W. 2012, \aap, 539, A18

\bibitem[{{Musiolik}(2021)}]{Musiolik2021}
{Musiolik}, G. 2021, \mnras, 506, 5153

\bibitem[{{Mutter} {et~al.}(2017){Mutter}, {Pierens}, \& {Nelson}}]{Mutter2017}
{Mutter}, M.~M., {Pierens}, A., \& {Nelson}, R.~P. 2017, \mnras, 465, 4735

\bibitem[{{Nelson} \& {Marzari}(2016)}]{NelsonMarzari2016}
{Nelson}, A.~F. \& {Marzari}, F. 2016, \apj, 827, 93

\bibitem[{{Neuh{\"a}user} {et~al.}(2007){Neuh{\"a}user}, {Mugrauer},
  {Fukagawa}, {Torres}, \& {Schmidt}}]{Neuhauser2007}
{Neuh{\"a}user}, R., {Mugrauer}, M., {Fukagawa}, M., {Torres}, G., \&
  {Schmidt}, T. 2007, \aap, 462, 777

\bibitem[{{Nigioni} {et~al.}(2026){Nigioni}, {Venturini}, {Bolmont}, {Turrini},
  {Alibert}, \& {Emsenhuber}}]{Nigioni2026}
{Nigioni}, A., {Venturini}, J., {Bolmont}, E., {et~al.} 2026, \aap, 708, A38

\bibitem[{{Offner} {et~al.}(2023){Offner}, {Moe}, {Kratter}, {Sadavoy},
  {Jensen}, \& {Tobin}}]{Offner2023}
{Offner}, S.~S.~R., {Moe}, M., {Kratter}, K.~M., {et~al.} 2023, in Astronomical
  Society of the Pacific Conference Series, Vol. 534, Protostars and Planets
  VII, ed. S.~{Inutsuka}, Y.~{Aikawa}, T.~{Muto}, K.~{Tomida}, \& M.~{Tamura},
  275

\bibitem[{{Owen} {et~al.}(2012){Owen}, {Clarke}, \& {Ercolano}}]{Owen2012}
{Owen}, J.~E., {Clarke}, C.~J., \& {Ercolano}, B. 2012, \mnras, 422, 1880

\bibitem[{{Paardekooper} {et~al.}(2008){Paardekooper}, {Th{\'e}bault}, \&
  {Mellema}}]{Paardekooper2008}
{Paardekooper}, S.-J., {Th{\'e}bault}, P., \& {Mellema}, G. 2008, \mnras, 386,
  973

\bibitem[{{Papaloizou} \& {Pringle}(1977)}]{PapaloizouPringle1977}
{Papaloizou}, J. \& {Pringle}, J.~E. 1977, \mnras, 181, 441

\bibitem[{{Pringle}(1981)}]{Pringle1981}
{Pringle}, J.~E. 1981, \araa, 19, 137

\bibitem[{{Raghavan} {et~al.}(2010){Raghavan}, {McAlister}, {Henry}, {Latham},
  {Marcy}, {Mason}, {Gies}, {White}, \& {ten Brummelaar}}]{Raghavan2010}
{Raghavan}, D., {McAlister}, H.~A., {Henry}, T.~J., {et~al.} 2010, \apjs, 190,
  1

\bibitem[{{Ragusa} {et~al.}(2016){Ragusa}, {Lodato}, \& {Price}}]{Ragusa2016}
{Ragusa}, E., {Lodato}, G., \& {Price}, D.~J. 2016, \mnras, 460, 1243

\bibitem[{{Ronco} {et~al.}(2021){Ronco}, {Guilera}, {Cuadra}, {Miller
  Bertolami}, {Cuello}, {Fontecilla}, {Poblete}, \& {Bayo}}]{Ronco2021}
{Ronco}, M.~P., {Guilera}, O.~M., {Cuadra}, J., {et~al.} 2021, \apj, 916, 113

\bibitem[{{Ronco} {et~al.}(2017){Ronco}, {Guilera}, \& {de
  El{\'{\i}}a}}]{Ronco2017}
{Ronco}, M.~P., {Guilera}, O.~M., \& {de El{\'{\i}}a}, G.~C. 2017, \mnras, 471,
  2753

\bibitem[{{Rosotti} \& {Clarke}(2018)}]{RosottiClarke2018}
{Rosotti}, G.~P. \& {Clarke}, C.~J. 2018, \mnras, 473, 5630

\bibitem[{{Rosotti} {et~al.}(2019){Rosotti}, {Tazzari}, {Booth}, {Testi},
  {Lodato}, \& {Clarke}}]{Rosotti2019b}
{Rosotti}, G.~P., {Tazzari}, M., {Booth}, R.~A., {et~al.} 2019, \mnras, 486,
  4829

\bibitem[{{Sellek} {et~al.}(2024){Sellek}, {Grassi}, {Picogna}, {Rab},
  {Clarke}, \& {Ercolano}}]{Sellek2024}
{Sellek}, A.~D., {Grassi}, T., {Picogna}, G., {et~al.} 2024, \aap, 690, A296

\bibitem[{{Shadmehri} {et~al.}(2018){Shadmehri}, {Ghoreyshi}, \&
  {Alipour}}]{Shadmehri2018}
{Shadmehri}, M., {Ghoreyshi}, S.~M., \& {Alipour}, N. 2018, \apj, 867, 41

\bibitem[{{Shakura} \& {Sunyaev}(1973)}]{Shakura1973}
{Shakura}, N.~I. \& {Sunyaev}, R.~A. 1973, \aap, 24, 337

\bibitem[{{Sullivan} {et~al.}(2023){Sullivan}, {Kraus}, {Huber}, {Petigura},
  {Evans}, {Dupuy}, {Zhang}, {Berger}, {Gaidos}, \& {Mann}}]{Sullivan2023}
{Sullivan}, K., {Kraus}, A.~L., {Huber}, D., {et~al.} 2023, \aj, 165, 177

\bibitem[{{Tanaka} \& {Ward}(2004)}]{Tanaka2004}
{Tanaka}, H. \& {Ward}, W.~R. 2004, \apj, 602, 388

\bibitem[{{Tang} {et~al.}(2017){Tang}, {MacFadyen}, \& {Haiman}}]{Tang2017}
{Tang}, Y., {MacFadyen}, A., \& {Haiman}, Z. 2017, \mnras, 469, 4258

\bibitem[{{Tanigawa} \& {Ikoma}(2007)}]{TanigawaIkoma2007}
{Tanigawa}, T. \& {Ikoma}, M. 2007, \apj, 667, 557

\bibitem[{{Tazzari} \& {Lodato}(2015)}]{TazzariLodato2015}
{Tazzari}, M. \& {Lodato}, G. 2015, \mnras, 449, 1118

\bibitem[{{Thebault} \& {Bonanni}(2025)}]{ThebaultBonanni2025}
{Thebault}, P. \& {Bonanni}, D. 2025, \aap, 700, A106

\bibitem[{{Th{\'e}bault} {et~al.}(2004){Th{\'e}bault}, {Marzari}, {Scholl},
  {Turrini}, \& {Barbieri}}]{Thebault2004}
{Th{\'e}bault}, P., {Marzari}, F., {Scholl}, H., {Turrini}, D., \& {Barbieri},
  M. 2004, \aap, 427, 1097

\bibitem[{{Vartanyan} {et~al.}(2016){Vartanyan}, {Garmilla}, \&
  {Rafikov}}]{Vartanyan2016}
{Vartanyan}, D., {Garmilla}, J.~A., \& {Rafikov}, R.~R. 2016, \apj, 816, 94

\bibitem[{{Venturini} {et~al.}(2020{\natexlab{a}}){Venturini}, {Guilera},
  {Haldemann}, {Ronco}, \& {Mordasini}}]{Venturini2020Letter}
{Venturini}, J., {Guilera}, O.~M., {Haldemann}, J., {Ronco}, M.~P., \&
  {Mordasini}, C. 2020{\natexlab{a}}, \aap, 643, L1

\bibitem[{{Venturini} {et~al.}(2020{\natexlab{b}}){Venturini}, {Guilera},
  {Ronco}, \& {Mordasini}}]{Venturini2020SE}
{Venturini}, J., {Guilera}, O.~M., {Ronco}, M.~P., \& {Mordasini}, C.
  2020{\natexlab{b}}, \aap, 644, A174

\bibitem[{{Venturini} {et~al.}(2026){Venturini}, {Nigioni}, {Ronco}, {Jungo},
  \& {Emsenhuber}}]{Venturini2026}
{Venturini}, J., {Nigioni}, A., {Ronco}, M.~P., {Jungo}, N., \& {Emsenhuber},
  A. 2026, \aap, 708, A37

\bibitem[{{Venturini} {et~al.}(2024){Venturini}, {Ronco}, {Guilera},
  {Haldemann}, {Mordasini}, \& {Miller Bertolami}}]{Venturini2024}
{Venturini}, J., {Ronco}, M.~P., {Guilera}, O.~M., {et~al.} 2024, \aap, 686, L9

\bibitem[{{Winter} \& {Haworth}(2022)}]{WinterHaworth2022}
{Winter}, A.~J. \& {Haworth}, T.~J. 2022, European Physical Journal Plus, 137,
  1132

\bibitem[{{Xie} \& {Zhou}(2008)}]{Xie2008}
{Xie}, J.-W. \& {Zhou}, J.-L. 2008, \apj, 686, 570

\bibitem[{{Youdin} \& {Goodman}(2005)}]{Youdin2005}
{Youdin}, A.~N. \& {Goodman}, J. 2005, \apj, 620, 459

\bibitem[{{Youdin} \& {Lithwick}(2007)}]{YoudinLithwick2007}
{Youdin}, A.~N. \& {Lithwick}, Y. 2007, \icarus, 192, 588

\bibitem[{{Zagaria} {et~al.}(2021){Zagaria}, {Rosotti}, \&
  {Lodato}}]{Zagaria2021}
{Zagaria}, F., {Rosotti}, G.~P., \& {Lodato}, G. 2021, \mnras, 504, 2235

\bibitem[{{Zurlo} {et~al.}(2023){Zurlo}, {Gratton}, {P{\'e}rez}, \&
  {Cieza}}]{Zurlo2023}
{Zurlo}, A., {Gratton}, R., {P{\'e}rez}, S., \& {Cieza}, L. 2023, European
  Physical Journal Plus, 138, 411

\end{thebibliography}

\begin{appendix}

\section{Dependence on $v_{\text{frag}}$}
\label{App:Vfrag}

The efficiency of pebble accretion depends sensitively on the size of the pebbles \citep{Lambrechts2012,LambrechtsJohansen2014} and therefore on the adopted fragmentation threshold velocity $v_{\text{frag}}$. In dust evolution models, this parameter controls the maximum size that particles can reach before destructive collisions occur, and thus strongly affects both the pebble flux and the pebble accretion rate.

Laboratory experiments and dust growth studies suggest typical values of $v_{\text{frag}}=1$~m~s$^{-1}$ for silicate grains and $v_{\text{frag}}\sim 5-10$~m~s$^{-1}$ for icy aggregates \citep[e.g.]
[]{GundlachBlum2015,Musiolik2021}. These values are commonly adopted in protoplanetary disc and planet formation models \citep{Drazkowska2016,Venturini2020Letter,Venturini2024}. However, the exact values remain uncertain, as they depend on particle composition, porosity, and the presence of volatile ices.

Our fiducial simulations consider for simplicity $v_{\text{frag}}=10$~m~s$^{-1}$ along the disc, as in \citet{Rosotti2019b,Zagaria2021}. To evaluate the sensitivity of our results to this parameter we performed two additional simulations adopting different fragmentation velocities along the disc, for the circular case. 

\begin{figure}[ht!]
  \centering
    \includegraphics[angle=0, width= \columnwidth]{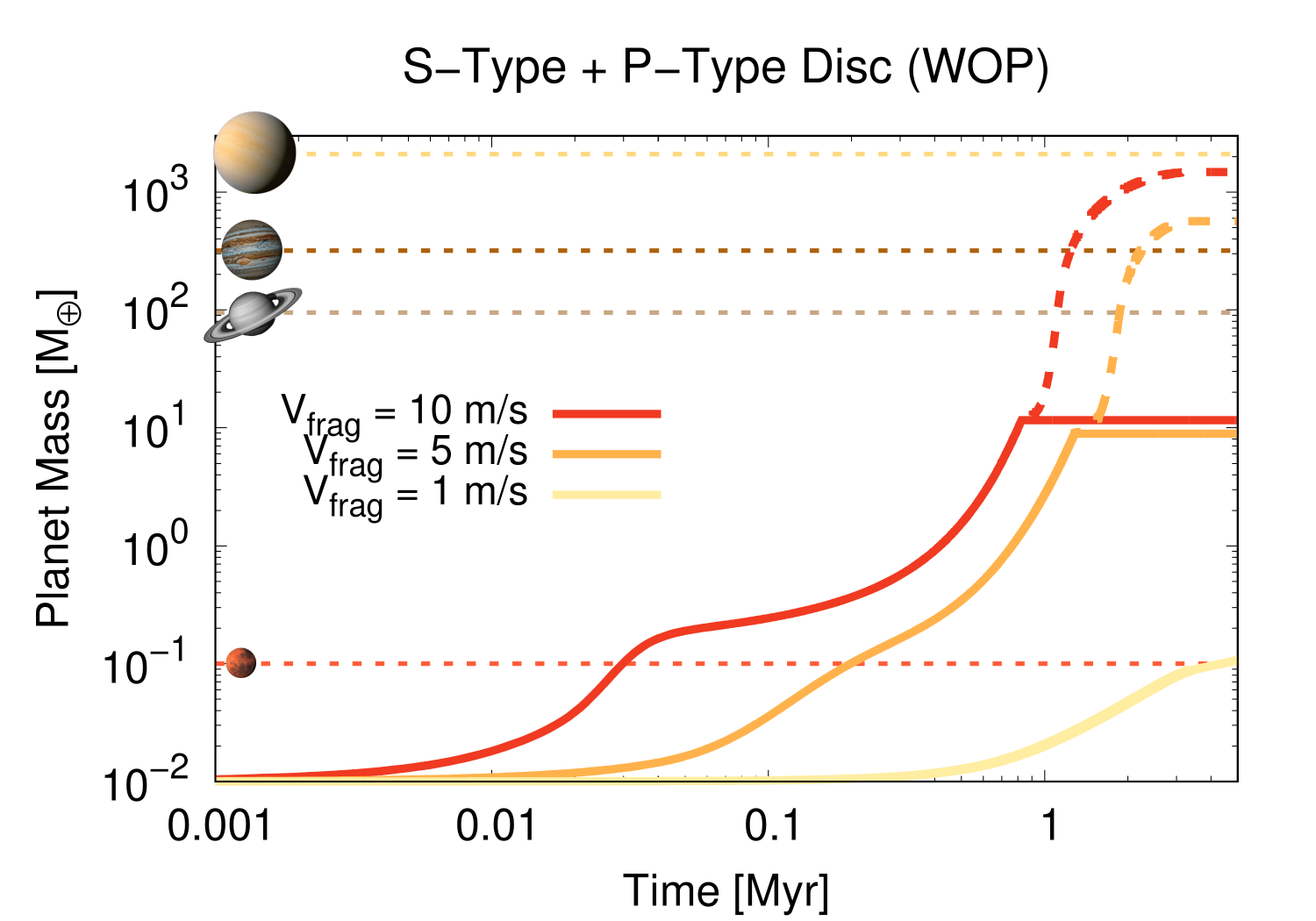}
    \caption{Planet formation tracks computed with three different values of the fragmentation velocity threshold of 10, 5 and 1 m s$^{-1}$, for a planet at 2 ua and only for the WOP case.}
\label{Formacion_Planetaria_Vfrag}
\end{figure}

Figure \ref{Formacion_Planetaria_Vfrag} shows evolutionary tracks for a planet that grows in-situ at 2 au considering $v_{\text{frag}}=10$~m~s$^{-1}$ (yellow curve), $v_{\text{frag}}=5$~m~s$^{-1}$ (orange curves) and 10 m s$^{-1}$. The results for $v_{\text{frag}}= 10$ m s$^{-1}$ are those already described in section \ref{sec:in_situ}. For $v_{\text{frag}}= 5$ m s$^{-1}$, the planet reaches its pebble isolation mass and still has time to accrete significant amounts of gas to form a few Jupiters-mass planet. Despite we did not reach the mass of $\gamma$-Cephei Ab, the planet growth is similar to the nominal case (red curve). This result suggests that decreasing the fragmentation velocity of icy grains by a factor of two does not alter our main conclusions.

For the simulation with $v_{\text{frag}}= 1$ m s$^{-1}$, the pebbles remain so small along the disc, that planet formation by pebble accretion is very inefficient and the initial embryo barely reaches the mass of Mars. Thus, the lower the fragmentation velocity threshold, the smaller the mass of the planet that can be formed. Our results therefore favor relatively high fragmentation velocities.

\section{Dependence on $M^{\text{ini}}_{\text{sys}}$}
\label{App:Msyst}

Another important parameter that could affect our planet formation outcomes is the initial mass of the system $M^{\text{ini}}_{\text{sys}}$. As mentioned in section \ref{IC}, $M^{\text{ini}}_{\text{sys}}$ is taken as a 10\% of the mass of the binary, that is, $M^{\text{ini}}_{\text{sys}}=0.1(M_{\text{P}}+M_{\text{S}})$. 
Usually, adopting an initial disc mass of about 10\% of the stellar mass is a common assumption in planet formation models, using the total mass of the binary rather than the mass of the primary star alone, as done here, could lead to an overestimate of the gas mass of the truncated CP disc and thus, to an overestimate of its solid budget to form planets. Therefore, to assess whether giant planets can still form for lower initial masses, we explore cases in which $M^{\text{ini}}_{\text{sys}}$ corresponds to 6\% and 3\% of the total binary mass.

\begin{figure}[ht!]
  \centering
    \includegraphics[angle=0, width= \columnwidth]{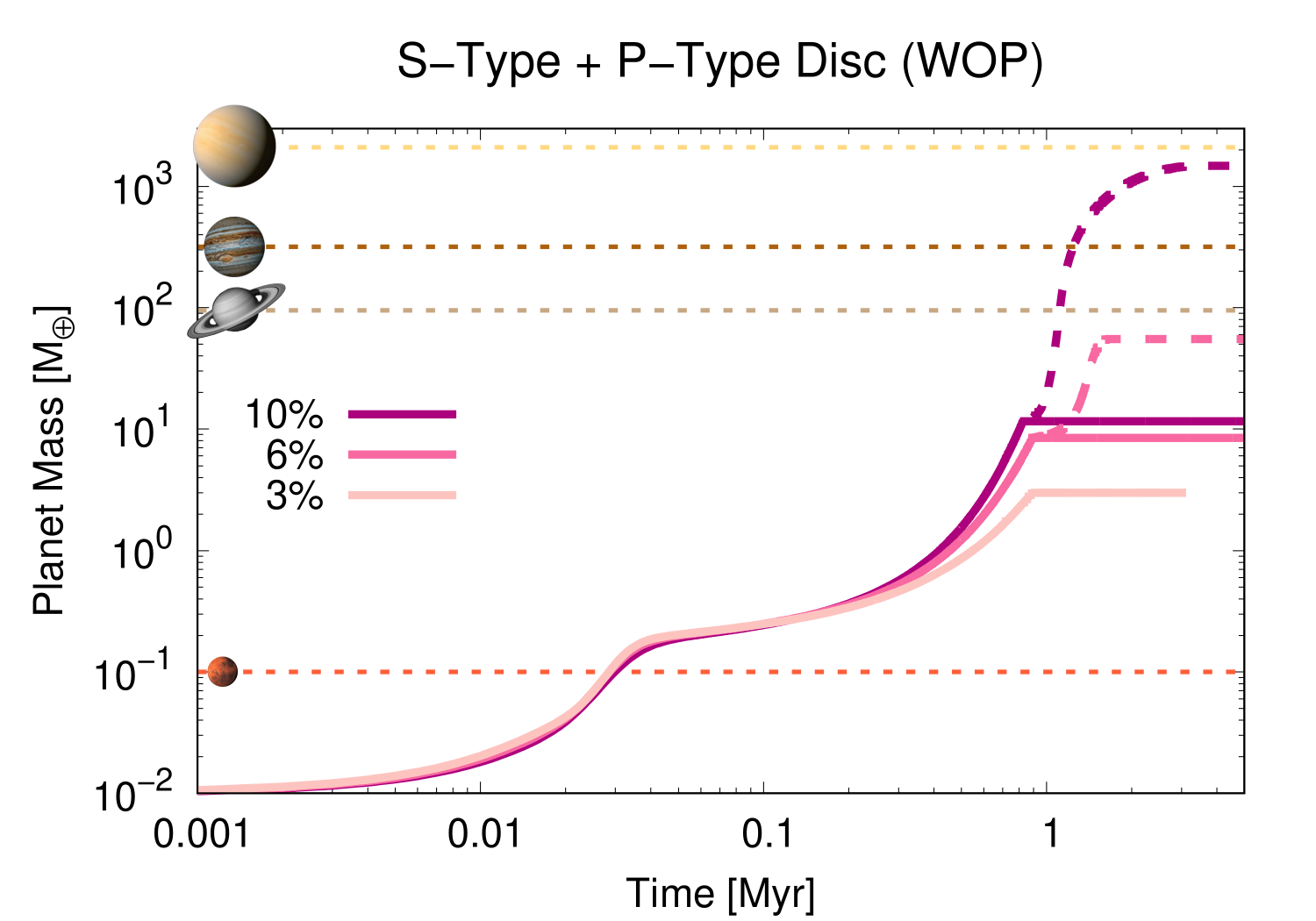}
    \caption{Planet formation tracks computed for three different percentages of the initial mass of the system, of 10\%, 6\% and 3\% of $M^{\text{tot}}_{\text{sys}}$, for a planet at 2 ua and only for the WOP case.}
\label{Formacion_Planetaria_Msist}
\end{figure}

Figure \ref{Formacion_Planetaria_Msist} presents a plot similar to Fig.~\ref{Formacion_Planetaria_Vfrag}, but for the different percentages of the binary mass. The results, only applied for the WOP case, clearly show that initial masses below 6\% of $M^{\text{ini}}_{\text{sys}}$ do not form enough massive planets. For the 3\% case, the planet at 2 au barely reaches a few Earth masses and in this case it does not reach its pebble isolation mass. In contrast, the planet in the 6\% case is able to almost form a Saturn-mass planet. Our results, as expected, strongly depend on the initial mass of the system, and the tendency is that the formation of gas giants like $\gamma$-Cephei Ab, achieved with the 10\% case alredady described in section \ref{sec:in_situ} is favored towards massive discs.

\section{Evolution of the CS disc}
\label{App:Disco_CS}

The results of \citet{MarzariDangelo2025} suggest that the CS disc around the lower-mass component of $\gamma$-Cephei likely did not receive significant amounts of gas or solids from the outer CB disc, nor through mass exchange with the CP disc. For this reason, in this work we focus primarily on the evolution of the replenished CP disc.

Regarding the CB disc, we consider two limiting scenarios: one in which its evolution is affected by X-ray photoevaporation from the inner binary (WP), and another in which this effect is neglected (WOP). Since the efficiency of this mechanism is uncertain, we speculate that the actual evolution of the system lies somewhere between these two extreme cases.

One possible factor that could reduce the impact of X-ray photoevaporation on the CB disc is the shielding provided by the circumstellar discs around each star. However, if the CS disc dissipates rapidly, as expected in the absence of replenishment from the CB reservoir, this shielding could weaken, potentially allowing photoevaporation of the CB disc to become more effective. This effect might be limited by the relatively low mass of the secondary star, which would likely drive a weaker photoevaporative wind. Nevertheless, if photoevaporation were partially mitigated by the presence of both circumstellar discs, the early dispersal of the CS disc could enhance the efficiency of CB disc dispersal. In that case, the most favorable conditions for planet formation could plausibly lie between the WOP and WP scenarios explored in this work.

To explore this possibility, we model the isolated evolution of the CS disc, considering both the circular and eccentric configurations of the system.

\begin{figure}[h]
  \centering
    \includegraphics[angle=0, width=1\linewidth]{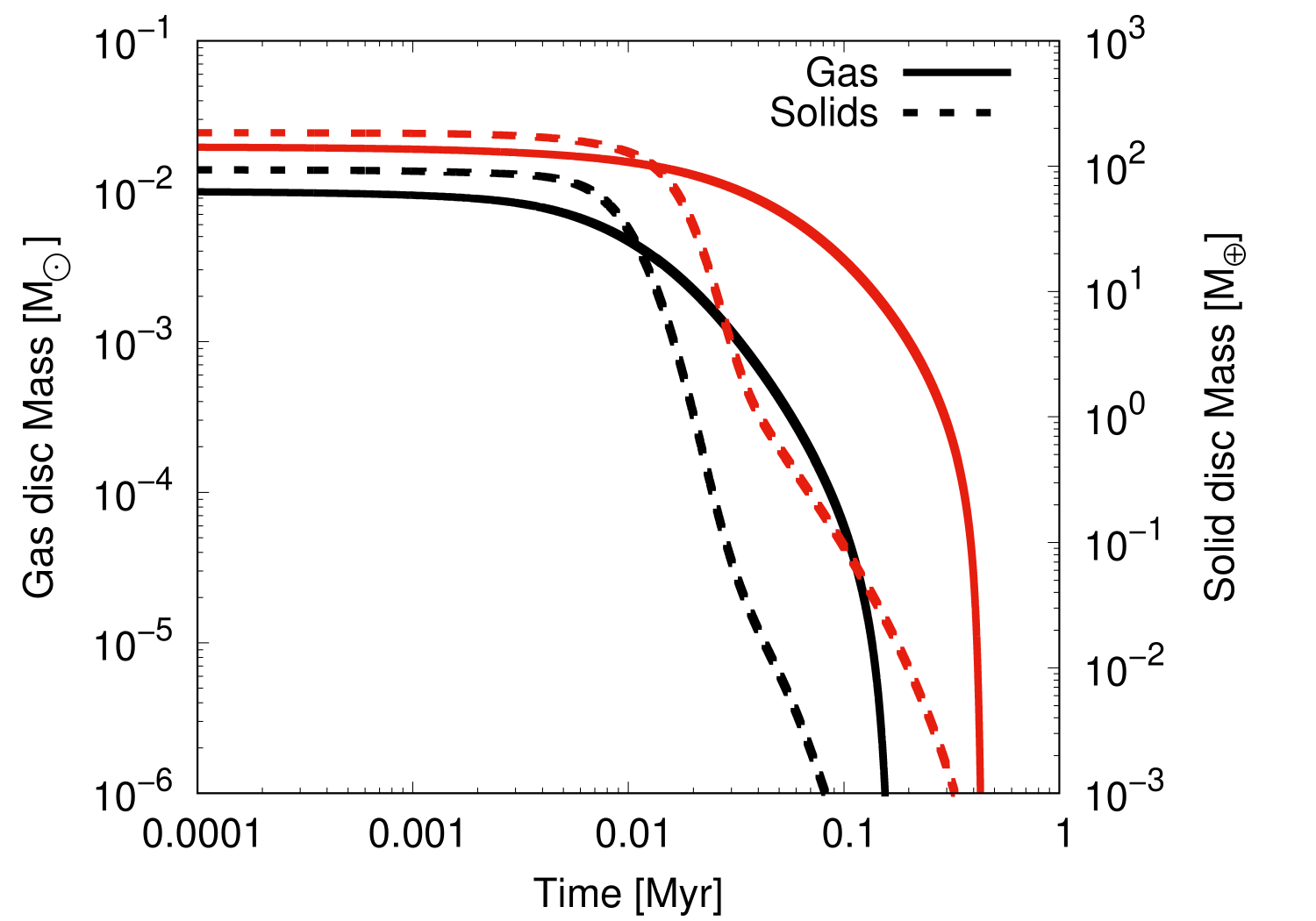}
    \caption{Mass of gas (solid lines, left label) and solids (dashed lines, right label) for the CS disc in the circular fiducial (red curves) and eccentric (black curves) cases.}
  \label{Disco_CS}
\end{figure}

Figure~\ref{Disco_CS} shows the evolution of the gas (solid lines) and dust (dashed lines) mass of the CS disc for the circular (red curves) and eccentric (black curves) cases. In the circular configuration, the gaseous disc dissipates after $\sim0.43$ Myr, while in the eccentric case it survives for less than $\sim0.2$ Myr, mainly due to its lower initial mass. The solid component evolves even more rapidly. In the circular case, only $\sim0.1,M_\oplus$ in pebbles remains after 0.1 Myr, whereas in the eccentric case the pebble mass at the same time is only $\sim0.1,M_{\rm Moon}$. These short dispersal timescales suggest that the CS disc would lose its ability to shield the CB disc relatively early, potentially allowing X-ray photoevaporation from the inner binary to become more effective. As a secondary remark, we note that with such small remaining solid reservoirs, and in the absence of replenishment from the CB disc, planet formation around the secondary star would not be expected.

\section{Eccentricity evolution of $\gamma-$Cephei-b ?}
\label{App:Nbody}
In this work, we have assumed that the growing planet remains in-situ and on a circular orbit throughout its formation. While these simplifications allows us to isolate the effects of disc evolution and mass replenishment on the planet growth, they are likely unrealistic, particularly in close binary systems such as $\gamma$~Cephei.

In the framework of secular dynamics, the planet's eccentricity can be described as the combination of a proper and a forced component, the latter being imposed by the binary companion. This behaviour has been analytically described in the restricted three-body framework by \citet{Heppenheimer1978}. Later, using a second-order perturbation theory, \citet{Giuppone2011} showed that the forced eccentricity can be well approximated by
\begin{equation}
e_f = \frac{5}{4}\frac{a_{\text{p}}}{a_{\text{B}}}\frac{e_{\text{B}}}{1-e_{\text{B}}^2}
\left[ 1 - 16\frac{M_{\text{P}}}{M_{\text{S}}}\frac{a_{\text{p}}^2}{a_{\text{B}}^2}(1-e_{\text{B}}^2)^{-5} \right],
\end{equation}
where $M_{\text{P}}$ and $M_{\text{S}}$ are the stellar masses of the primary and secondary, respectively, $a_{\text{p}}$ and $a_{\text{B}}$ are the semimajor axes of the planet and the secondary star, respectively, and $e_\text{B}$ is the binary eccentricity. Thus, as a consequence of the purely gravitational perturbation from an eccentric secondary, the long-term evolution of the planetary eccentricity is characterized by oscillations about the forced value, with amplitudes that are set by the initial conditions. 

Dissipative effects, such as the interaction with a disc, tend to damp the proper eccentricity, but cannot remove the forced component. As a result, the planet is expected to evolve towards a state in which its eccentricity oscillates around a non-zero forced value set by the binary configuration.

To validate this theoretical expectation in our problem, we perform N-body numerical simulations of the CP planet in the $\gamma$-Cephei system, inmersed in two different CP discs. For the gravitational interaction we consider the restricted problem in which the planet does not perturb the binary orbit. For the disc interaction, we assume that dissipation only affects the planetary orbit by damping the eccentricity, in timescales characterized by $\tau_{\text{e}}$. Following the Type-I migration recipe of \citet{Tanaka2004}, this timescale can be estimated from the disc property as: 
\begin{equation}
\tau_{\text{e}} = \frac{1}{0.78} \frac{M_{\text{P}}}{m_{\text{p}}} \frac{M_{\text{P}}}{\Sigma^{\text{CP}}_{\text{g}}(r)r^2} \frac{h(r)^4}{\Omega_K)} 
\end{equation}
where $m_{\text{p}}$ represents the planetary mass, $\Omega_K$ the local Keplerian frequency, $\Sigma^{\text{CP}}_{\text{g}}(r)$ the CP disc surface density at radius $r$, and $h(r)$ the corresponding disc aspect ratio.

In our N-body simulations, we set the planet at a fixed $a = 2 \mathrm{au}$, which is very close to the current position of $\gamma$-Cephei Ab. The eccentricity damping timescale $\tau_{\text{e}}$ is computed at each instant of time as the disc evolves for this fixed distance. Then, the perturbation of such a damping is modeled through an ad-hoc Stokes-type force \citep[e.g.,][]{Beauge2006}, added to the pure gravitational interaction within the binary system.

\begin{figure}[t]
\includegraphics[width=1\linewidth]{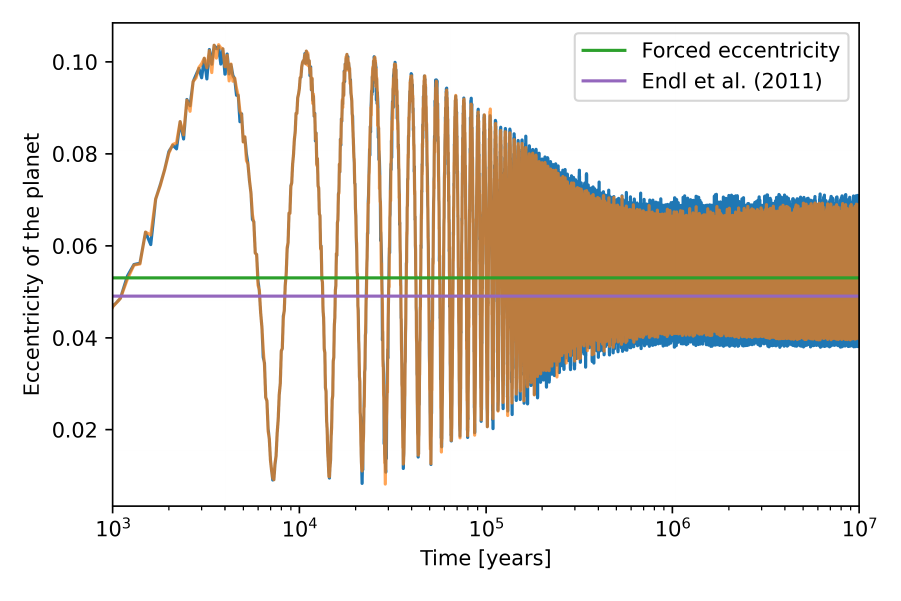}
\caption{ N-body simulation of the eccentricity evolution of the CP planet growing within a CP disc in the $\gamma$-Cephei system. The blue curve represents the evolution of the planet in the isolated disc, without gas replenishment from the outer CB disc, and the orange curve represents the evolution in the WOP case. The horizontal green line corresponds to the analytical fit of the forced eccentricity obtained by \cite{Giuppone2011}. The magenta horizontal line represents the mean value of the eccentricity solution of \cite{Endl2011}.}
\label{Nbody-App}
\end{figure}

Figure \ref{Nbody-App} shows the N-body eccentricity evolution of the CP planet within the isolated disc (blue) and the WOP disc (orange) cases. The green horizontal line represents the analytical fit of the forced eccentricity obtained \cite{Giuppone2011} while the magenta horizontal line corresponds to the orbital solution obtained by \cite{Endl2011} with a compilation of radial velocity data.

We note that the \cite{Endl2011} eccentricity is very close to the mean value of the eccentricity of the planet in our simulation. Such proximity to the equilibrium is an indicator that some dissipative mechanism should have played a role in the orbital evolution of the planet. Then, from \ref{Nbody-App} we note that the numerical simulations reproduce very well the theoretical expectation: in both cases, as the proper eccentricity is being damped due to the dissipation,  the eccentricity evolves toward the equilibrium characterized by oscillations around the forced value, which is very well fitted by the analytical model. The only small difference between disc models is observed in the oscillation amplitudes, being slightly smaller in the WOP case. Thus, as expected from the secular theoretical models and also from the numerical simulations performed here, the early formation and dynamical evolution of the forming planet in $\gamma$-Cephei probably was in an eccentric orbit. This state is forced by the pure gravitational interaction with the binary (instead of a single star) and can not be avoided by any dissipation related to the formation process. 

The exact mean value of this quantity depends on the physical and orbital parameters of the system. However, if the initial binary orbit was similar to the one observed today, this value depends primarily on the semimajor axis at which the planet formed and evolved. Assuming that the planet did not undergo significant radial migration, we estimate that its mean eccentricity during the early stages of evolution was of order $10^{-2}$. According to \cite{Liu2018}, pebble accretion onto protoplanets with eccentricities of this magnitude can substantially enhance the accretion efficiency, potentially increasing the final planetary mass by a factor of a few. Therefore, in addition to the presence of a CB protoplanetary disc in the $\gamma$-Cephei system, accounting for the planet's eccentricity evolution may also be important as an additional key ingredient for the formation of a giant planet in such a hostile environment.

\begin{figure}[h]
  \centering
    
\end{figure}

\end{appendix}

\end{document}